\documentclass[usenatbib]{aa}  
\usepackage{graphicx}
\usepackage{txfonts}
\usepackage[pdfpagelabels=false]{hyperref}	
\hypersetup{colorlinks=true,linkcolor=blue,citecolor=blue,filecolor=blue,urlcolor=blue}

\begin{document}

   \title{Extinction and AGN over host galaxy contrast effects on the optical spectroscopic classification of AGN}

   \subtitle{}

   \author{L. Barquín-González
          \inst{\ref{IFCA}}
          \and
          S. Mateos
          \inst{\ref{IFCA}}
          \and
          F. J. Carrera
          \inst{\ref{IFCA}}
          \and
          I. Ordovás-Pascual
          \inst{\ref{IFCA}}
          \and
          A. Alonso-Herrero
          \inst{\ref{CAB}}
          \and
          A. Caccianiga
          \inst{\ref{BRERA}}
          \and
          N. Cardiel
          \inst{\ref{COMPLU}}
          \and
          A. Corral
          \inst{\ref{IFCA}}
          \and
          R.M. Domínguez
          \inst{\ref{IFCA}}
          \and
          I. García-Bernete
          \inst{\ref{OXFORD}}
          \and
          G. Mountrichas
          \inst{\ref{IFCA}}
          \and
          P. Severgnini
          \inst{\ref{BRERA}}
          }

   \institute{
   	IFCA (University of Cantabria-CSIC), Avenida de los Castros, 39005 Santander, Spain\label{IFCA}\\
    \email{barquin@ifca.unican.es}
    \and
    Centro de Astrobiología (CAB), CSIC-INTA, Camino Bajo del Castillo s/n, 28692 Villanueva de la Cañada, Madrid, Spain\label{CAB}
    \and
    INAF - Osservatorio Astronomico di Brera, via Brera 28, I-20121 Milan, Italy\label{BRERA}
    \and
    Departamento de Física de la Tierra y Astrofísica, Fac. CC. Físicas, Universidad Complutense de Madrid, Plaza de las Ciencias 1, E-28040, Spain\label{COMPLU}
    \and
    Astrophysics, University of Oxford, DWB, Keble Road, Oxford OX1 3RH, UK \label{OXFORD}
	}

	\date{April 29, 2024}

  \abstract{The optical spectroscopic classification of active galactic nuclei (AGN) into type 1 and type 2 can be understood in the frame of the AGN unification models. However, it remains unclear which physical properties are driving the classification into intermediate sub-types (1.0,1.2,1.5,1.8,1.9). To shed light on this issue, we present an analysis of the effect of extinction and AGN and host galaxy luminosities on sub-type determination for a sample of 159 X-ray selected AGN with a complete and robust optical spectroscopic classification. The sample spans a rest-frame 2 -- 10 keV X-ray luminosity range of $10^{42}$--$10^{46}$ erg s$^{-1}$ and redshifts between 0.05 and 0.75. From the fitting of their UV-to-mid-infrared spectral energy distributions, we extracted the observed AGN over total AGN+galaxy contrast, optical/UV line-of-sight extinction as well as host galaxy and AGN luminosities. The observed contrast exhibits a clear decline with sub-type, distinguishing two main groups: 1.0-5 and 1.8-9/2. This difference is partly driven by an increase in extinction following the same trend. Nevertheless, 50$\%$ of 1.9s and 2s lack sufficient extinction to explain the lack of detection of broad emission lines, unveiling the necessity of an additional effect. Our findings show that 1.8-9/2s preferentially live in host galaxies with higher luminosities while displaying similar intrinsic AGN luminosities to 1.0-5s. Consequently, the AGN to host galaxy luminosity ratio diminishes, hindering the detection of the emission of the broad emission lines, resulting in the 1.8-9/2 classification of those with insufficient extinction. Thus, the combination of increasing extinction and decreasing AGN/galaxy luminosity ratio, mainly driven by an increasing host galaxy luminosity, constitutes the main reasons behind the sub-type classification into 1.0-5 and 1.8-9/2.}
  

	\titlerunning{Extinction and luminosity effects on AGN optical spectroscopial classification}
	\authorrunning{L. Barquín-González et al.}
   \maketitle
%
\nolinenumbers
\section{Introduction}
\label{sec:introduction}
	According to standard unification models of active galactic nuclei (AGN), their observed optical spectral properties can mostly be explained as the effect of anisotropic obscuration of their nuclear emission \citep{antonucci1993, urry1995,netzer2015}. This obscuration arises from a toroidal distribution of gas and dust, commonly referred to as the 'torus' \citep{gallimore2016,imanishi2018,garciaburillo2016,garciaburillo2019,garciaburillo2021,alonsoherrero2018}. In the traditional AGN models, the inclination of the line of sight with respect to the axis perpendicular to the equatorial plane of the torus is a fundamental parameter, categorizing the AGN population into two main groups: type 1 and type 2. AGN are classified as type 1 when the inclination is preferentially low. In this population, we have a direct vision of the nuclear region and its characteristic emission at optical wavelengths: broad permitted emission lines (FWHM>$1000$ km s$^{-1}$) from the broad line region (BLR) and a prominent blue continuum from the accretion disk. On the other hand, AGN are classified as type 2 if they are observed at high inclinations, close to the equatorial plane of the torus, when the emission from the central engine is absorbed. For these no blue continuum or broad permitted emission lines are detected, instead, we will be able only to detect narrow emission lines (FWHM<$1000$ km s$^{-1}$), both forbidden and permitted, also present in type 1. Recent observations in the infrared range showed the torus is best described with a clumpy distribution of the obscuring material \citep{herrero2003,elitzur2006,nenkova2008a,nenkova2008b,Markowitz2014}. In these models detection of nuclear emission depends on the probability of the line-of-sight being intercepted by a dusty cloud, allowing for the detection of type 1 AGN with high inclination and type 2 AGN with low inclination \citep{elitzur2012}.
	
	After the identification of type 1 and type 2 AGN, \cite{osterbrock1977} demonstrated that there is a broad distribution of flux ratios between the broad and narrow components of H$\beta$ among type 1s. This revealed that some AGN can be understood as intermediate steps between a pure type 1 or sub-type 1.0, with a very strong broad component of H$\beta$ dominating the total H$\beta$ profile, and a type 2, for which no broad component is detected. Subsequently, sub-types 1.0, 1.2, 1.5 and 1.8 were defined using the flux ratio of a broad and narrow emission line, usually H$\beta$ and [OIII]5007, and the sub-type 1.9 as those AGN without detection of a broad component for H$\beta$ but with the detection of a broad component for H$\alpha$ \citep{osterbrock1981,winkler1992,Whittle1992}. Intermediate type AGN are a typical occurrence both at low redshift \citep{Kyuseok2022,mejia-restrepo} and high redshift, where recent studies with the James Webb Space Telescope (JWST) have shown intermediate type AGN are the majority of newly discovered AGN \citep{mathee2023,harikane2023,kocevski2023,maiolino2023}. However, it is still neither clear what is the physical origin of the different AGN sub-types nor if the same mechanism is responsible for the classification in the different redshift regimes.
	
	
	A straightforward interpolation between type 1 and type 2 AGN might imply that sub-types are linked to an increasing obscuration. Previous works have indeed identified an increase in optical extinction and X-ray absorption with classification \citep{herrero2003,Schnorr2016,burtscher2016}. However, further studies also revealed a population of unobscured type 1.8-9, which directly challenges the extinction scenario \citep{Caccianiga2004,trippe2010,zhang2023}. 
	
	Other differences between type 1s and type 2s may be interpolated in the same way as obscuration in order to explain the emergence of the intermediate classification. It has been shown in the literature that type 2s are found in host galaxies with higher stellar mass compared to type 1s \citep{zou2019,mountrichas2021,koutoudilis2022}. This could be related to higher host galaxy luminosity, or to an increase of host galaxy dust, and consequently, optical extinction associated with the interstellar medium \citep{malkan1998,Guainazzi2005}. Furthermore, 1.8/1.9/2s AGN are more probably found in edge-on galaxies than 1.0/1.2/1.5 \cite{maiolino1995,winter2009,koss2011}. Another factor contributing to the change in sub-type may be differences in AGN intrinsic luminosities. Several surveys at optical, infrared and X-ray wavelengths have shown that type 2 AGN have preferably lower luminosities than type 1 AGN \citep{dellaceca2008,ueda2014,Koss2017,Suh2019}. If the same phenomenon is happening for the different sub-types, it would make AGN observed features less intense, explaining the different classification.
	
	In this study, we aim to shed light on what physical properties play a main role in the optical spectroscopic classification, focussing on identifying the effect that both optical extinction  and AGN and host galaxy luminosities have on the sub-type classification. To this end, we fitted the rest-frame UV/optical spectroscopic observations and rest-frame UV-to-mid-infrared spectral energy distributions of an X-ray selected AGN sample. 
	The AGN sample used in this work has been drawn from the Bright Ultra-hard XMM-Newton Survey (BUXS; \citealt{mateos2012,mateos2015}), a sample of bright, luminous X-ray detected AGN. We have an almost complete (96\%) and uniform optical spectroscopic classification up to $z=0.75$, based on the detection of broad emission lines and on the flux ratios of [OIII]$\lambda$5007 and the broad component of H$\beta$. Thanks to the high spectroscopic identification completeness and the use of uniform classification criteria, our study will not suffer from biases related to classification incompleteness, more severe for sub-types 1.9 and 2 as they are fainter and more difficult to determine class and redshift. This makes our dataset ideally suited for the study of the dependence of AGN subtype with AGN and host galaxy properties. The large sample size ensures that we can extract robust statistical conclusions from the comparison between sub-types.
	
	
	This paper is organized as follows. Sect. \ref{sec:2} introduces the parent sample and the selection of our working sample. In Sect. \ref{sec:3}, we describe the spectral and photometric fitting process, as well as the classification scheme. In Sect. \ref{sec:4} we present and discuss our main result. Conclusions are reported in Sect. \ref{sec:5}. In this paper, we will use the reddening, $E(B-V)$, as a proxy of the optical extinction, $A_V$, and use both terms indistinctly as they are directly correlated by $R_V$. Throughout this paper, errors are at 1$\sigma$ confidence level and we use a cosmology with parameters $\Omega_\text{M}=0.3$, $\Omega_\Lambda=0.7$ and $H_\text{0}=70$ km s$^{-1}$ Mpc$^{-1}$.


	\section{Sample}
	\label{sec:2}
	
	The AGN sample used in this study has been drawn from the Bright Ultra-hard XMM-Newton Survey (BUXS)\footnote{\href{https://buxs.unican.es/}{https://buxs.unican.es/}}. BUXS is a complete sample of 255 X-ray bright AGN detected with the \textit{XMM-Newton} European Photon Imaging Camera (EPIC)-pn \citep{struder2001} and with 4.5 -- 10 keV fluxes over $6\times10^{-14}$ erg s$^{-1}$ cm$^{-2}$. The selected range of energies allows for reducing the bias against highly absorbed AGN. For all the AGN in this work, good quality \textit{XMM-Newton} spectra were already available in the observed energy range from 0.25 to 10 keV. The median number of counts after the background subtraction was 1437. A careful modelling of the X-ray spectra has been carried out to derive X-ray luminosities in the rest-frame 2 -- 10 keV range, $L_X$. From this point, everytime we will talk about $L_X$ we will refer to the decimal logarithm of its value in ergs per second. Full details of this analysis will be presented in Mateos et al. (in prep.). A complete description of the parent sample, including a table with all the objects forming it, can be found in \cite{mateos2012,mateos2015}.
	
\begin{figure}
		\centering
		\includegraphics[width=\linewidth]{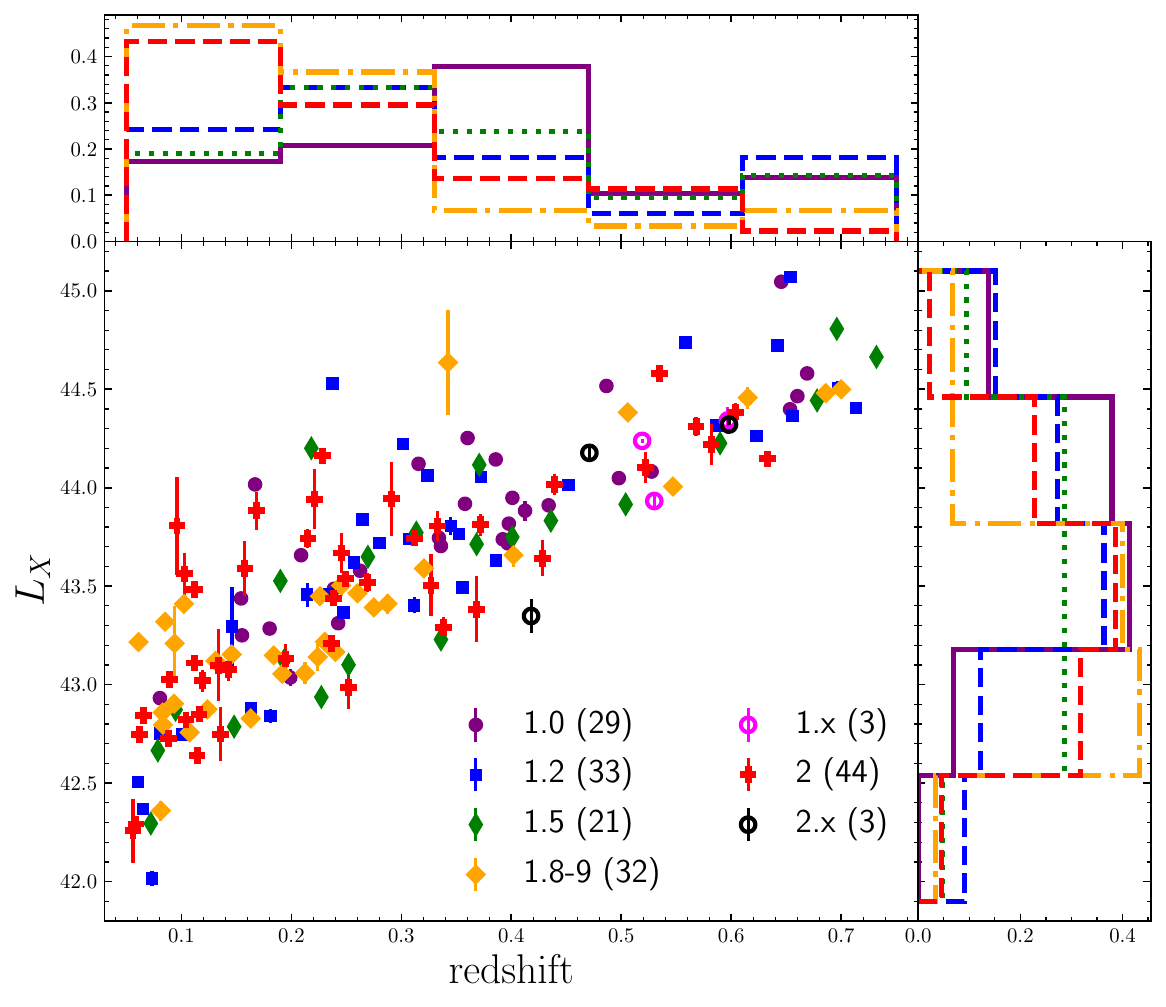}
		\caption{Decimal logarithm of the intrinsic rest-frame 2-10 keV X-ray luminosity versus redshift, and the projected distributions of these quantities for different sub-types of AGN (see text). Distributions are normalised such that the sum of the heights equals 1. Symbols, line styles and colour codes of this figure depend on sub-type (see Sect. \ref{sec:classification}) and will remain the same for the rest of the work.}
		\label{fig:lx_vs_z}
	\end{figure}

	In this work, we have selected those AGN with redshift lower than $z=0.75$ to ensure we have a high identification completeness, 96\%, as for higher redshifts H$\beta$ leaves the coverage range of our optical spectra. A description of the follow-up spectroscopic observations and the classification used in this work will be introduced in Sects. \ref{sec:spectra} and \ref{sec:classification}, respectively. As customary in other works based on X-ray selected samples, we have also introduced a cut-off of $L_X=10^{42}$ erg s$^{-1}$. The L$_X$-z distribution of the 165 AGN is illustrated in Fig. \ref{fig:lx_vs_z}.

\section{Spectral and photometric analysis}
\label{sec:3}

\subsection{Optical/NIR spectral analysis}
\label{sec:spectra}

\begin{table}
	\centering
	\caption{Summary of the properties of the optical/NIR spectra.} 
	\label{tab:instrumentation_spec}
	\begin{tabular}{cccc}
		\hline\hline
		Instrument & N & $\lambda$ range (\AA) & R $(\lambda/\Delta\lambda)$ \\
		(1) & (2) & (3) & (4) \\
		\hline
		GTC/OSIRIS   		& 12 & 4800 -- 10\,000	& 587  \\
		NOT/ALFOSC 			& 2  & 3200 -- 9700	& 360  \\
							& 1  & 3650 -- 7110 & 650  \\
		NTT/EFOSC2   		& 2  & 4085 -- 7520 & 370  \\
							& 1  & 3685 -- 9315 & 210  \\
		SDSS/SDSS-I/II   	& 71 & 3800 -- 9100 & 1850 -- 2200  \\
		SDSS/BOSS   		& 30 & 3600 -- 10\,400& 1560 -- 2560  \\
		SUBARU/FOCAS   		& 3  & 4700 -- 9100 & 500  \\
		TNG/DOLORES   		& 14 & 3000 -- 8430 & 585  \\
		UH88/WFGS			& 1  & 3800 -- 9700 & 440  \\
		VLT/FORS2   		& 12 & 4450 -- 8650 & 440  \\
		WHT/ACAM 			& 1  & 3300 -- 9500 & 570  \\
		WHT/ISIS			& 15 & 5300 -- 9300 & 844  \\ \hline
		GTC/EMIR			& 14 & 8500  -- 13\,500 & 987  \\
		WHT/LIRIS			& 1  & 9760 -- 13\,300 & 700  \\
		\hline
	\end{tabular}
	
	\tablefoot{Col. 1: name of the telescopes and instrumentes; Col. 2: number of spectra obtained with that instrument; Col. 3: wavelength coverage; Col. 4: resolution for a slit width of 1\farcs, except for the WFGS and EMIR which are given for a slit of 1.1\farcs. The two rows for NOT/ALFOSC and NTT/EFOSC2 correspond to different grism configurations. The last two rows correspond to NIR spectographs while the rest are for optical spectrographs.}
\end{table}

Spectral classifications, accurate redshifts and rest-frame UV/optical emission line intensities were determined through the fitting of the optical and/or NIR spectra in the regions around MgII$\lambdaup\lambdaup2796,2803 \AA$, H$\beta$ and H$\alpha$. To carry out the spectral analysis several Python scripts were developed using SHERPA's modeling and fitting applications as its framework \citep{sherpa,shepa_python}. The Levenberg-Marquardt optimization method \citep{levmar}, implemented by SHERPA, was employed to search for the local minimum in the fitting procedure. Confidence intervals for 1$\sigma$ were calculated using the \textit{conf} method. This method computes the confidence interval bounds for the model parameters varying one of them while allowing the others to float to new best-fit values. The fitting process for the MgII$\lambdaup\lambdaup2796,2803 \AA$, H$\beta$ and H$\alpha$ regions was performed independently. For fitting the continuum, we employed a local model consisting of an accretion disk, host galaxy and FeII components. The emission lines were modelled using multiple Gaussians. A comprehensive explanation of the spectral fitting procedure can be found in Appendix \ref{sec:apendix_emission_lines}.

We collected data for 101 AGN from the public catalogues of the Sloan Digital Sky Survey (SDSS) Data Release 17 \citep{dr17}. Among these, 30 observations were conducted with the BOSS Spectrograph and 71 were carried out with the old SDSS spectrograph. The remaining spectra were obtained in our own spectroscopic campaigns. The main properties of the optical/NIR spectrographs used are summarized in Table \ref{tab:instrumentation_spec}. 

With our redshift cut of $z=0.75$, we can assure coverage of H$\beta$ for all AGN. However, we start to lack coverage of H$\alpha$ in the optical range at $z\gtrsim0.4$. To ensure we use the same classification schema for all objects in the sample we carried out a follow-up at NIR wavelengths for 15 AGN without a broad H$\beta$ detection and for which H$\alpha$ lies outside the coverage of the optical spectra. After combining all spectral information, we have coverage of the region around H$\beta$ for $>99\%$ AGN in our sample and of H$\alpha$ for $133$ AGN ($>80\%$ of the sample).

Reduction of the non-SDSS data has been carried out using IRAF \citep{doug1986}. A full description of the reduction process is outside the scope of this work and will be presented in Mateos et al. (in prep.). The quality of our spectra is good, for 90\% we have $\mathrm{S/N}>5$\footnote{S/N has been calculated as the median value of the ratio between the observed flux density and its associated uncertainty among bins in the whole spectral range.}. We have corrected for the effect of the Milky Way extinction in all spectra using the model of \cite{allen1976} and the Galactic N$_H$ map of \cite{DL1990} with $R_V=3.1$. 

Thanks to the combination of all these observations, we have a sample with a homogeneous criterion for the classification of all sources up to $z\sim0.75$.

\subsection{AGN classification}
\label{sec:classification}

We classified our AGN using a scheme based on that defined by \cite{Whittle1992}. AGN are classified into sub-types 1.0, 1.2, 1.5 and 1.8 using the observed flux ratio between [OIII]$\lambdaup 5007$ and the broad component of H$\beta$, $R=F(\mathrm{[OIII]}\lambdaup5007)/F(\mathrm{H\beta^b})$, as follows: 

\begin{itemize}
	\item 1.0 when $R$$\leq0.3$;
	\item 1.2 when $0.3<R\leq1$;
	\item 1.5 when $1<R\leq 4$;
	\item 1.8 when $R>4$;
\end{itemize}

Besides these sub-types, AGN are classified as 1.9 when we detect a broad component for H$\alpha$ but not for H$\beta$. Finally, type 2 AGN are identified as those AGN where neither a broad component for H$\alpha$ nor for H$\beta$ are detected. This classification scheme may be affected by the projected aperture of the observations. We explored this in Appendix \ref{sec:proyected_aperture} and found no significant aperture effect in the class determination of our sample.

There are 6 AGN for which a reliable intermediate classification could not be determined. In two of those a broad component of MgII emission line was detected, but no coverage of H$\beta$ was available. In another, we detected a broad component of H$\beta$ but atmospheric absorption made it impossible to measure reliably the flux of [OIII]$\lambdaup 5007$. These three cases have been labelled as 1.x in Fig. \ref{fig:lx_vs_z}. There are also three AGN for which we did not detect a broad component of H$\beta$ and we did not have coverage of H$\alpha$. They were labelled as 2.x in Fig. \ref{fig:lx_vs_z}. All these 6 cases were removed from the sample. The remaining 159 AGN is what we will call onward the working sample from which we will derive our results. The number of AGN in each sub-type and their median $L_X$ and redshift can be found in Table \ref{tab:subtypes_stats}. For the rest of the work, we will consider the sub-types as numerical values. So increasing sub-type means going from 1.0 to 2 in the order described just above and a decreasing classification the opposite. 

\begin{figure}
	\centering
	\includegraphics[width=\columnwidth]{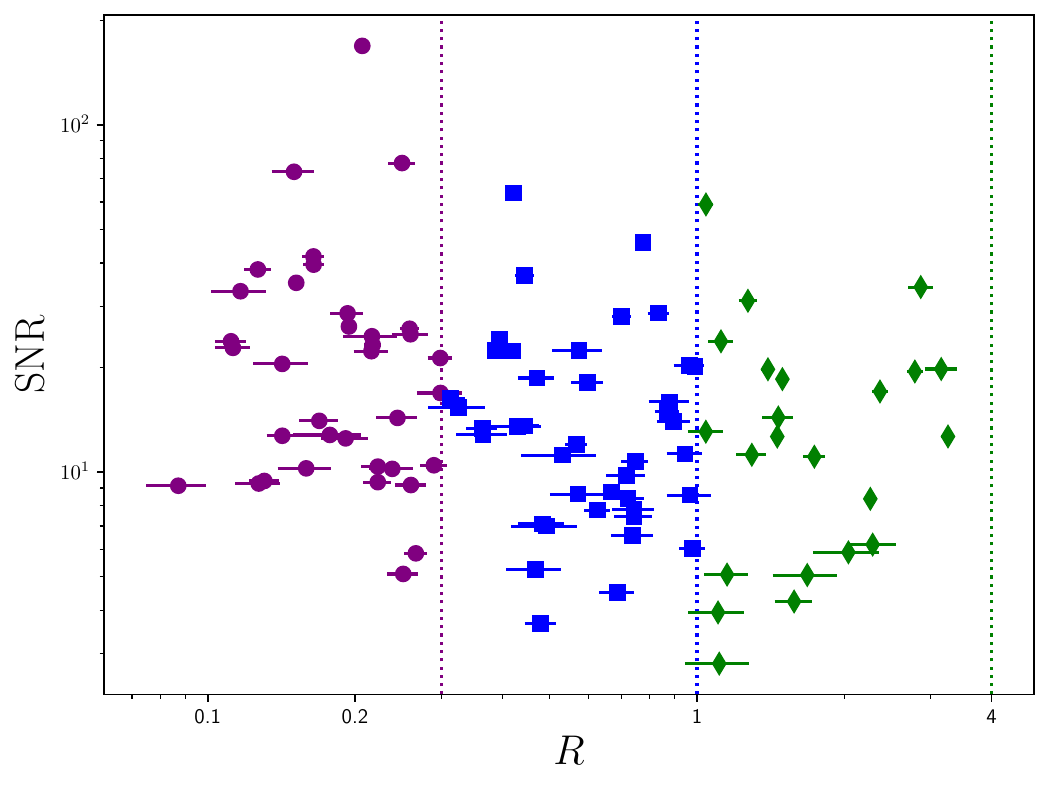}
	\caption{SNR of the 1.0 (purple dots), 1.2 (blue squares) and 1.5 (green diamonds) AGN samples versus R. The dotted vertical lines correspond to the limit values between one sub-type and the following. }
	\label{fig:R_Dist}
\end{figure}

We looked for possible cross-contamination between 1.0/1.2/1.5 due to the uncertainty in the determination of R. To determine if there was any important mix between sub-types we plotted $R$ against SNR, with threshold values superposed, in Fig. \ref{fig:R_Dist}. Any potential contamination should be confined to the few AGN close to the thresholds between sub-types, up to 10 AGN. For most however, it is highly unlikely due to the small uncertainty of $R$. Given also that there is no correlation between SNR and $f_{\mathrm{[OIII]}\lambdaup5007}/f_{\mathrm{H\beta^b}}$ we can also be sure the quality of the individual spectra is not playing a role in the determination of the sub-type.

The detection of broad H$\beta$ becomes increasingly difficult with R, especially in the spectra with the lowest S/N. Then, as long as we detect a broad H$\alpha$ we cannot discard the presence of a very weak undetected broad H$\beta$. To estimate the impact of the 1.5-8 contamination in our 1.9 AGN sample, we have checked those fits that initially included a broad component of H$\beta$ but it was below our detection significance threshold (5$\sigma$, see Appendix \ref{sec:apendix_emission_lines}). After a thoughtful visual check-up, we found a maximum of three objects with a genuine broad component of H$\beta$ below our significance threshold. This would translate in a variation in the 1.9 sample no higher than 10\%. As the sample of 1.8 AGN is too small (only 2 objects) to obtain any statistically significant result, we merged the 1.8s with the 1.9s. From now on, we will refer to the merged sample as 1.8-9.

We have also checked for the impact of 1.9s with low significance H$\alpha$ in the type 2 AGN sample. Following the same approach as before, we found the variation would be no higher than six objects (14\%).

\begin{table}
	\centering
	\caption{AGN properties of the different samples.} 
	\label{tab:subtypes_stats}
	\begin{tabular}{cccccc}
		\hline\hline
		Sample & Number & $\langle L_\mathrm{X} \rangle$ & $\sigma(L_\mathrm{X})$ &
		$\langle z \rangle$ & $\sigma(z)$\\
		(1) & (2) & (3) & (5) & (6)& (7) \\
		\hline
		1.0   & 29 & $43.88$ & 0.47 & 0.36 & 0.16\\
		1.2   & 33 & $43.74$ & 0.72 & 0.34 & 0.20\\
		1.5   & 21 & $43.71$ & 0.67 & 0.34 & 0.20\\
		1.8-9 & 32 & $43.22$ & 0.56 & 0.25 & 0.18\\
		2.0   & 44 & $43.51$ & 0.56 & 0.25 & 0.16\\
		\hline& & 
	\end{tabular}
	\tablefoot{Col. 1: sample by sub-type; Col. 2: total number of the sub-type sample; Cols. 3, 4: mean value and standard deviation of the rest-frame 2 -- 10 keV luminosity; Cols. 5, 6: mean value and standard deviation of the redshift.}
\end{table}	

\subsection{SED fitting}
\label{sec:SED}

We have fitted the rest-frame UV-to-mid-infrarred SEDs in order to obtain the optical/UV extinction and the continuum luminosity of both AGN and host galaxy at rest-frame 5100$\AA$, an usual reference wavelength in the literature, which is also close to H$\beta$.

Our SEDs are assembled with data from the Galaxy Evolution Explorer (GALEX; \citealt{bianchi2017}), Sloan Digital Sky Survey (SDSS; \citealt{Abazajian2009}), the Two Micron All Sky Survey (2MASS; \citealt{Jarret2000,Cutrit2003}), the UKIRT INfrared Deep Sky Survey (UKIDSS; \cite{lawrence2007}), the Visible and Infrared Survey Telescope for Astronomy (VISTA; \citealt{vista_a2006,vista_b2006}) and the Wide-Field Infrared Survey Explorer (WISE; \citealt{Wright2010}). All photometry data were corrected for the Milky Way extinction in the same way as the spectra. We added to the flux densities an additional 10\% uncertainty in quadrature to consider uncertainties in absolute flux calibration and any other potential unknown factors.

We have utilized the SED fitting program SEd Analysis using BAyesian Statistics (SEABAS; \citealt{rovilos2014}). It fits the SED with AGN and host galaxy templates using a Monte Carlo Markov chain (MCMC) sampling method. For modelling the AGN emission we used a combination of an accretion disk and a torus template. To model the torus emission we employed the Seyfert 1 template and the three Seyfert 2 templates from \cite{silva2004} corresponding to column densities $\log_{10}(\mathrm{N}_H/\mathrm{cm}^{-2})=\{ 22-23,23-24,24-25\}$ cm$^{-2}$. For the disk, we used the type 1 quasar SED from \cite{richards2006} up to $\lambdaup =0.7\mu$m. For longer wavelengths, we have extrapolated the disk emission with a power-law $\lambdaup f_\lambda\propto \lambdaup^{-1}$. We have fixed the intrinsic flux ratio of the unabsorbed disk ($E(B-V)=0$) and torus of an unobscured AGN ($\log_{10}(\mathrm{N}_H/\mathrm{cm}^{-2})=0$) at rest-frame 6 $\mathrm{\mu}$m, $C_{6\mu \text{m}}$, to the observed median value for sub-type 1.0, which typically have low $E(B-V)$. To calculate it we did a first fit of all type 1.0 without fixing $C_{6\mu \text{m}}$. From these fits we obtained the intrinsic value after correcting for the fitted extinction, obtaining $C_{6\mu \text{m}}=0.10\pm0.04$. We checked if imposing a fixed $C_{6\mu \text{m}}$ affects the extinction estimation of the 1.0s and found no significant changes. $C_{6\mu \text{m}}$ is equivalent to $L_{IR}/L_{BOL}$, commonly used in the literature, and our value is similar to that found in previous studies for all AGN sub-types  \citep{richards2006,lusso2013,roseboom2013,ichikawa2019}.

We have reddened the disk template with the Small Magellanic Cloud extinction law from \cite{gordon&clayton1998} for $\lambdaup<3000\AA$ and the Galactic extinction law from \cite{cardelli1989} for $\lambdaup>3000\AA$. In both cases, we have assumed a $R_V=3.1$. To redden the templates we have used a range of $E(B-V)$ from 0 to 2, with a grid of $\Delta E(B-V)=0.01$ up to 0.65 and a grid of $\Delta E(B-V)=0.15$ onwards. The increased grid for low values of $E(B-V)$ allows us to look for potential differences at low extinction values where the extinction difference effect is less prominent.  We allow for any combination of values of $E(B-V)$ and $\log_{10}(\mathrm{N}_H/\mathrm{cm}^{-2})$ to account for all the possible shapes of the AGN emission. The SEABASs application doesn't allow for the estimation of uncertainties in the extinction fitted. To estimate the significance of our extinction results with respect to certain threshold levels, we have forced a refit with that specific value of $E(B-V)$ and study $\Delta\chi^2$ (see below and Sect. \ref{sec:obscuration}).

Using a fixed disk/torus normalisation, we are able to have an estimate of the disk extinction, even for the highest values of our grid and even when the galaxy dominates, since the rest-frame MIR points of the SED constrain the torus and, from it, the intrinsic disk luminosity. The value of $E(B-V)$ estimated this way accounts for the combined contribution of nuclear and host galaxy-associated components, in the rest of the paper, this total optical/UV extinction in the line-of-sight towards the AGN will be referred to as simply extinction, unless otherwise stated explicitly.

\begin{figure}
	\centering
	\includegraphics[width=\columnwidth]{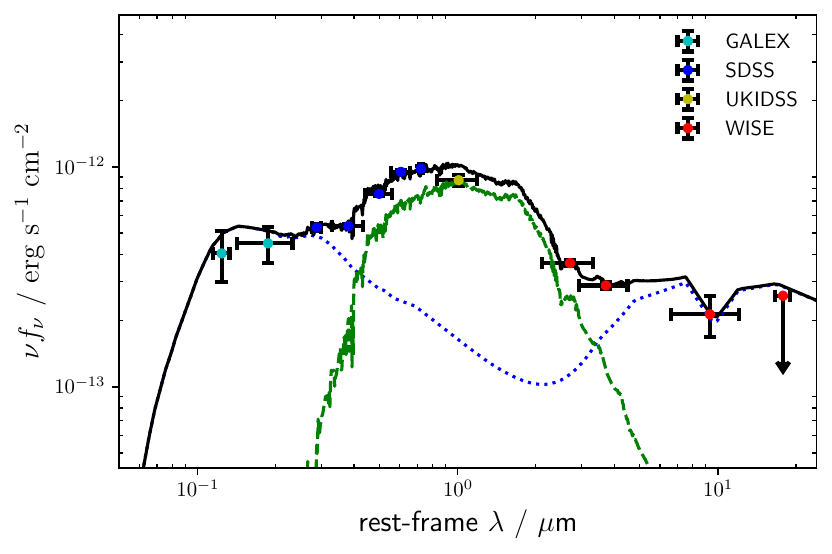}
	\includegraphics[width=\columnwidth]{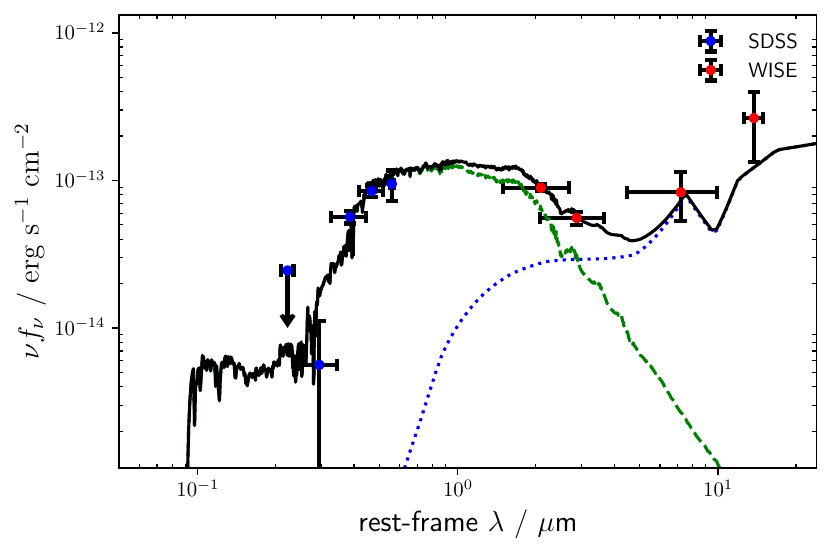}
	\caption{Two examples of the SED decomposition analysis. Coloured symbols corresponds to the different surveys the data has been taken from. The upper fit is correspond to a type 1.0 AGN. The blue dotted line corresponds to a reddened accretion disk with $E(B-V)=0.02$ plus torus template with N$_{\mathrm{H}}$=10$^{22}$ cm$^{-2}$, while the dashed-green line corresponds to the host galaxy component with $E(B-V)_{\mathrm{GAL}}=0.057$. The lower fit correspond to a type 2 AGN with $E(B-V)=2.0$, N$_{\mathrm{H}}$=10$^{24}$ cm$^{-2}$ and $E(B-V)_{\mathrm{GAL}}=0.062$.}
	\label{fig:SED_example}
\end{figure}

For modelling the host galaxy we have used a library of 75 stellar templates from \cite{bruzual2003}. Our templates have solar metallicity and a Chabrier initial mass function \citep{chabrier2003}. To generate them we have used 10 exponentially decaying star formation histories with characteristic times $\tau$ = 0.1–30 Gyr, a model with constant star formation, and a set of ages in the range 0.1–13 Gyr. The host galaxy templates were later reddened using the \cite{calzetti2000} dust extinction law and a range of $E(B-V)_{\mathrm{GAL}}$ from 0 to 2.  We have not added a contribution by dust heated by star formation to avoid potential degeneracies between AGN and host galaxy templates. We can safely do this as shown in \cite{mateos2015}, where conservative upper-limits for the star formation contamination at $6\ \mu m$ were obtained for the BUXS sample. For more than 90\% of the objects with $L_X>10^{42}\ \mathrm{erg\ s^{-1}}$, contamination was estimated to be below 15\%, having negligible effect on our results. Examples of two fitted SEDs, a type 1.0 and a type 2, are shown in Fig. \ref{fig:SED_example}.

To determine which best fits are compatible with no extinction at all, we have refitted all objects imposing $E(B-V)=0$. If the significance of the improvement is less than 90\% ($\Delta\chi^2<2.71$) the value fitted would be identified as compatible with no extinction. Additionally, if the extinction is comparable to the uncertainty of the Galactic correction we also mark these fits as compatible with no extinction. We have found that $70\%$ of the objects with a non-zero value of E(B-V) and less or equal to $0.05$ are compatible with no obscuration at all. We keep the fitted values for the sake of obtaining the best possible fit, but mark them in all plots as compatible with no extinction with a triangle pointing down.

For some cases, the best fit SED resulted in a very bright and blue galaxy to describe what should be fitted with an unobscured disk at rest-frame UV wavelengths. There is a known degeneracy between fitting a very blue disk without extinction with a faint host galaxy, or a bright blue host galaxy with an absorbed disk at these wavelengths as their shapes are similar \citep{Suh2019,marshall2022}. To identify those cases, we compare the SED fit with the optical spectrum. If after a visual analysis of the optical spectra we conclude the continuum is clearly dominated by the disk emission, we limit the maximum extinction of the disk to $E(B-V)$=0.1 to favour the fit of an unobscured disk. This was necessary for 13 objects (8\%): six Type 1.0, six Type 1.2 and one Type 1.5 AGN.

\begin{figure}
	\centering
	\includegraphics[width=\columnwidth]{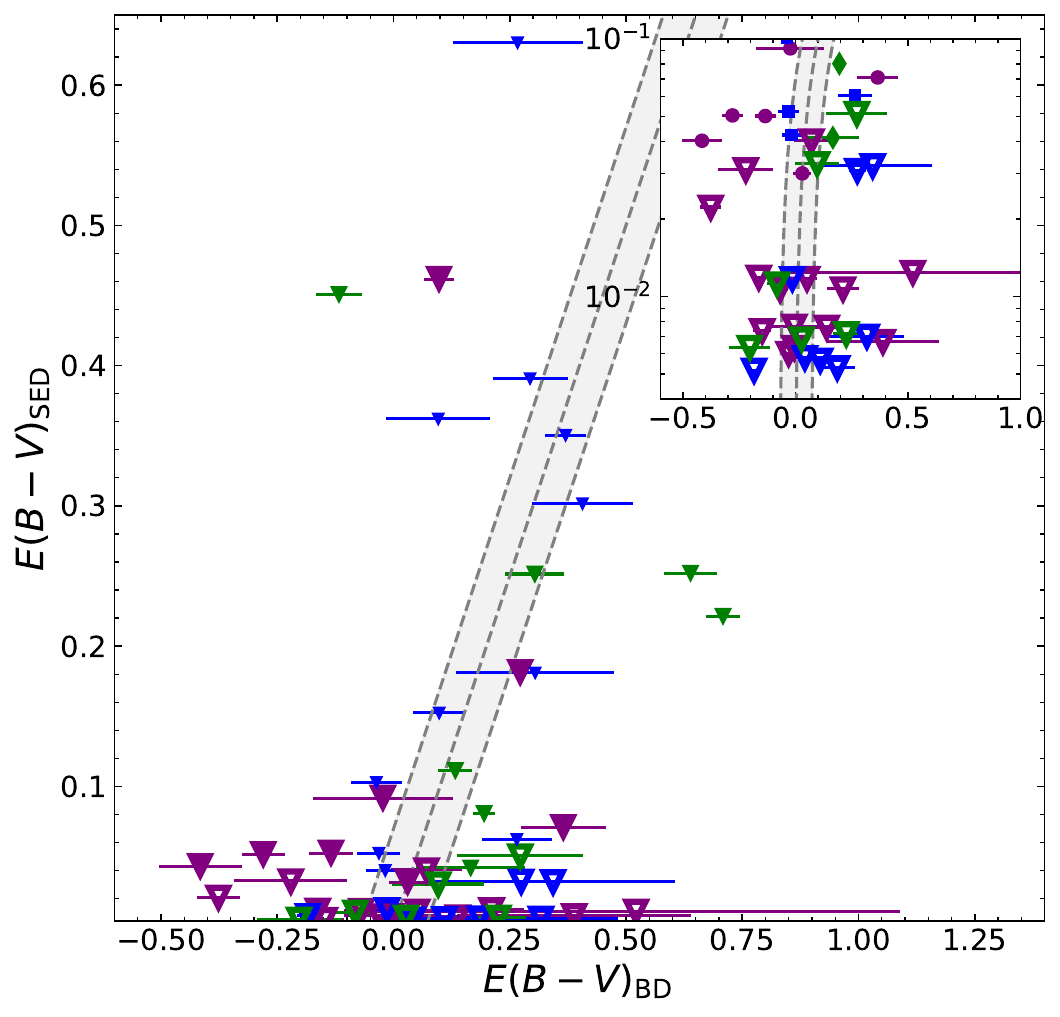}
	\caption{Optical extinction of the accretion disk derived from the SED fit ($E(B-V)_{\mathrm{AGN,SED}}$) versus the one derived from the Balmer decrement of the H$\alpha$ and H$\beta$ broad emission lines ($E(B-V)_{\mathrm{AGN,BD}}$). The grey area represents the 1:1 relation and its 1$\sigma$ uncertainty from \protect\cite{dong2008}. A random contribution of $\pm0.003$ has been added to ${E(B-V)_{\text{SED}}}$ to avoid the superposition of points with the same value. The inset shows a zoom of those objects with ${E(B-V)_{\text{SED}}<0.1}$. Symbol and colour coded as in Fig. \ref{fig:lx_vs_z}, except for AGN with E(B-V) compatible with no obscuration, that have been represented with empty triangles pointing down.}
	\label{fig:BD_LineasSED}
\end{figure}

Another way to obtain a value of extinction is from the difference between the observed and intrinsic Balmer decrement of the broad emission lines. In this work, we have chosen to determine extinctions using the SED instead because estimates based on Balmer decrements have important limitations. First, it can only be applied to AGN with both detected broad H$\beta$ and H$\alpha$, i.e., sub-types 1.0, 1.2, 1.5 and 1.8, which does not allow us to study the whole sample but simply a subset. Secondly, it depends heavily on the still debated intrinsic line flux ratio \citep{dong2008,gaskell2017,lu2019}. However, as Balmer decrement-based extinctions are frequently  used in the literature, we checked if they correlated well with those derived from the SED fits. We obtained the Balmer decrements for those 1.0-5s with broad H$\beta$ and H$\alpha$ detected and coming from the same spectra. To convert it to an extinction value, we use an intrinsic ratio of $F(\mathrm{H}\alpha^{\text{b}})/F(\mathrm{H}\beta^{\text{b}})=3.06\pm0.03$ \citep{dong2008}. In Fig. \ref{fig:BD_LineasSED} we can see how the two estimates of the optical extinction are well correlated. From this point any reference to $E(B-V)$ will refer to the values obtained from the SED fits.

\begin{table*}
	\centering
	\caption{KS test p-values results for the null hypothesis that the samples under consideration have the same distribution for different parameters of the AGN or host galaxy.} 
	\label{tab:kstest}
	\begin{tabular}{cccccccc} 
		\hline \hline
		Sample 1 & Sample 2 & $C_{obs}$ &
		$E(B-V)$& $\mathrm{log}_{10}[F$(H$\alpha^\text{n}$)/$F$(H$\beta^\text{n}$)] & $L_{\text{AGN}}$ & $L_{\text{GAL}}$ & $L_{\text{AGN}}/L_{\text{GAL}}$ \\
		(1) & (2) & (3) & (4) & (5) & (6) & (7) & (8)\\
		\hline
		1.0 & 1.2   & \textit{0.006} 		& 0.117   		 & {0.383}  & {0.074}   	 & {0.113}   	  & \textit{0.029}\\
		1.2 & 1.5   & {0.506}        		& {0.534}   	 & {0.637}  & {0.815}   	 & {0.245}   	  & {0.462}\\
		1.5 & 1.8-9 & \textbf{0.000} 		& \textbf{0.000} & {0.422}  & {0.855}		 & \textbf{0.000} & \textit{0.005}\\
		1.8-9 & 2.0 & {0.086}        		& {0.607}   	 & {0.055}  & {0.184}   	 & {0.072}   	  & {0.880}\\
		1.0 & 1.5   & \textit{0.009} 		& \textit{0.014} & {0.331}  & \textit{0.012} & {0.339}		  & \textit{0.020}\\
		1.2 & 1.8-9 & \textbf{0.000}		& \textbf{0.000} & {0.458}  & {0.939}   	 & {0.056}    	  & \textit{0.009}\\
		1.5 & 2.0   & \textbf{0.000} 		& \textbf{0.000} & {0.086}  & {0.918}		 & \textbf{0.000}   	  & \textit{0.005}\\
		1.0-5 & 1.8-9/2.0 & \textbf{0.000}	& \textbf{0.000} & {0.856}  & {0.062}		 & \textbf{0.001}  	  & \textbf{0.000}\\
		1.0-5$_{\mathrm{Low\ }E(B-V)}$ & 1.8-2.0$_{\mathrm{Low\ }E(B-V)}$ & ...  & ...  & ...  & \textit{0.009}    & \textbf{0.000} & \textbf{0.000}\\ \hline
		
	\end{tabular}
	\tablefoot{We consider a result as significant if it is greater than 3$\sigma$ ($P(H_0)<0.003$). In this case, we mark the result in bold. If the result lies between 3$\sigma$ and 2$\sigma$ ($0.003<P(H_0)<0.05$) we display it in italic. Cols. 1, 2: samples evaluated by the KS test; Col. 3: observed contrast of the AGN over the total flux at rest-frame 5100\AA; Col 4: extinction of the AGN emission; Col 5: Balmer decrement of the narrow emission lines; Col 6: host-galaxy luminosity at rest-frame 5100\AA \ weighted by redshift distributions (see Sect. \ref{sec:lum}); Col 7: absorption-corrected luminosity of the AGN at rest-frame 5100\AA\ weighted by redshift distributions (see Sect. \ref{sec:lum}); Col 8: ratio between host galaxy and AGN optical luminosity at rest-frame 5100\AA.}
\end{table*}

	\section{Results and discussion}
	\label{sec:4}
	
	In this section, we will show the main results of the comparison of several parameters between intermediate classes and discuss their implications. Below, when we compare two different distributions we will refer to the result of the Kolmogorov--Smirnov (KS) test to statistically quantify their differences. We will identify a difference as significant if the probability of the null hypothesis\footnote{The null hypothesis assumes that both distributions has been drawn from the same parent distribution.}, $P(H_0)$, is lower than 0.003, i.e., it is greater than $3\sigma$. The P-values of every KS test run in this work are shown in Table \ref{tab:kstest}. We also repeat the comparison for Cols. 3 -- 5 and 8 with the Anderson-Darling test, which is more sensitive to tails, finding similar results for all tests shown here. The same could not be done for Cols. 6 and 7 as we used a special KS test implementation for these distributions to take into account differences in the redshift distributions of the samples (see Sect. \ref{sec:lum}).
	
	\subsection{Observed contrast}
	\label{sec:contrast}
	
\begin{figure}
		\centering
		\includegraphics[width=\columnwidth]{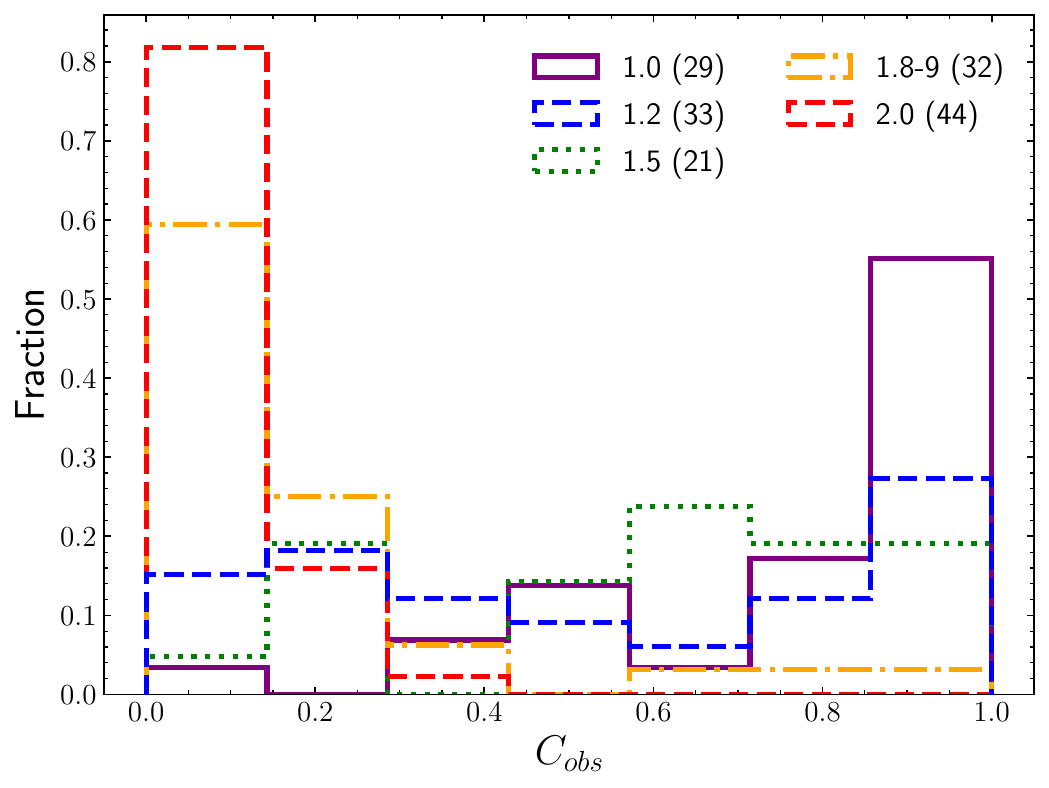}
		\caption{Observed contrast of the AGN over the total flux at rest-frame $5100\AA$. Line styles as in Fig. \ref{fig:lx_vs_z}.}
		\label{fig:contrast}
	\end{figure}

	In this work we have defined the observed contrast as,
	
	\begin{equation}
		\label{ec:contrast}
		C_\mathrm{obs}=f_{\text{AGN}}/(f_{\text{AGN}}+f_{\text{GAL}}).
	\end{equation}

	where $f_{\text{AGN}}$ and $f_{\text{GAL}}$ are the observed AGN and host galaxy flux density at rest-frame 5100\AA, respectively. Our first step was to calculate the observed contrast of every sub-type and compare them. The values of $f_{\text{AGN}}$ and $f_{\text{GAL}}$ were obtained from the SED best-fit templates. The distribution of the observed contrast is shown in Fig. \ref{fig:contrast}. The KS results indicate that there are two significantly different groups: one composed of 1.0/1.2/1.5 and another of 1.8-9/2.0 AGN, with the former having an overall higher contrast than the latter. No significant differences are found between individual sub-types within these groups; although it is hinted a difference between 1.0s and 1.2s ($P(H_0)=0.006$) and 1.0s and 1.5s ($P(H_0)=0.009$). Encouraged by these results, we will conduct any subsequent comparison also for the two main groups identified (hereafter referenced as 1.0-5 and 1.8-9/2) for the rest of this work.
	
	To properly explain the change in class we need to identify which physical parameters, or which combination of them, leads to this decrease of contrast with increasing sub-type. This could be due to an increase on the extinction level or to changes in the intrinsic luminosity of the host galaxy or AGN. The first physical parameter we checked for potential differences is the extinction.

	\subsection{AGN and host galaxy extinction}
	
	\label{sec:obscuration}
	
\begin{figure}
		\centering
		\includegraphics[width=\columnwidth]{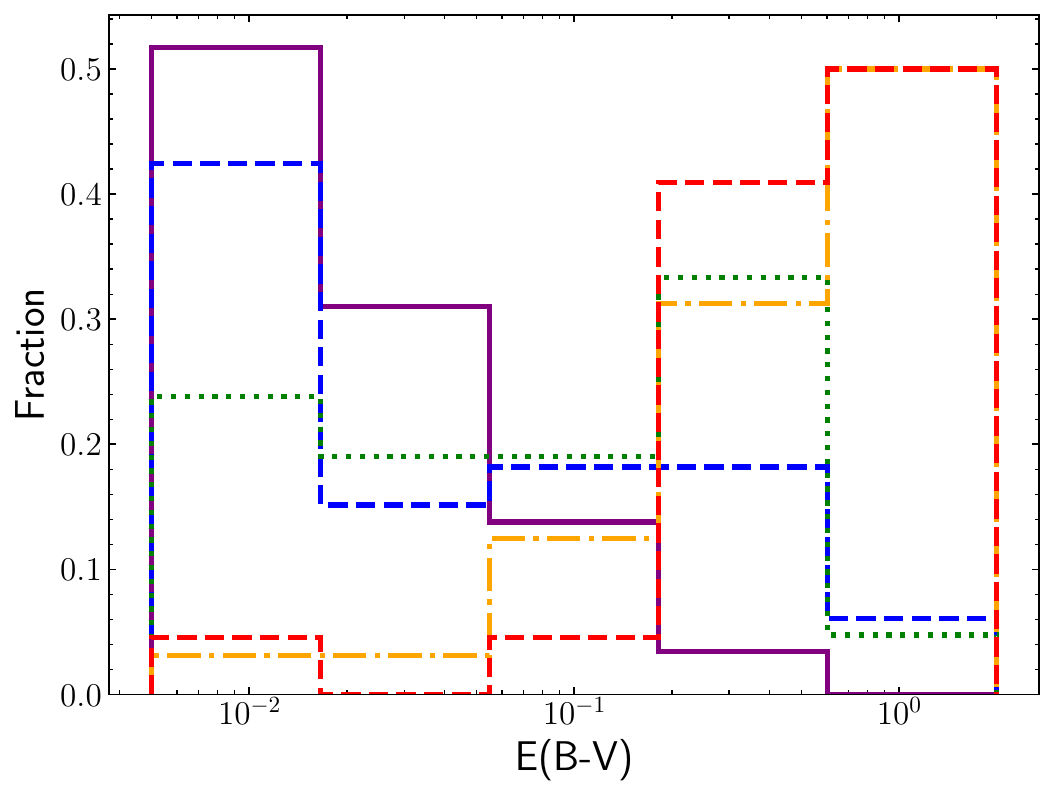}
		\caption{Line-of-sight optical extinction of the AGN accretion disk component. Values of $E(B-V)=0$ have been added to the lowest bin. The appearance of the distributions and KS results don't change significantly if $E(B-V)$ values compatible with no obscuration are treated as $E(B-V)=0$. Line styles as in Fig. \ref{ec:contrast}.}
		\label{fig:EBV_dist}
	\end{figure}

	Changes in the amount of extinction may explain the decrease in contrast observed. To identify if this is the situation, we compared the distributions of extinction in intermediate types in Fig. \ref{fig:EBV_dist}. According to the results of the KS test (see Table \ref{tab:kstest}), we found that 1.0-5 AGN are significantly less extinguished than 1.8-9/2 AGN, while no significant differences are found between finer sub-types. Furthermore, the difference in observed contrast previously hinted between 1.0s and 1.2s becomes less relevant ($P(H_0)=0.117$). Consequently, even if 1.0s have higher contrast, this is unlikely to be due to their lower overall extinction.
	
\begin{figure}
		\centering
		\includegraphics[width=\columnwidth]{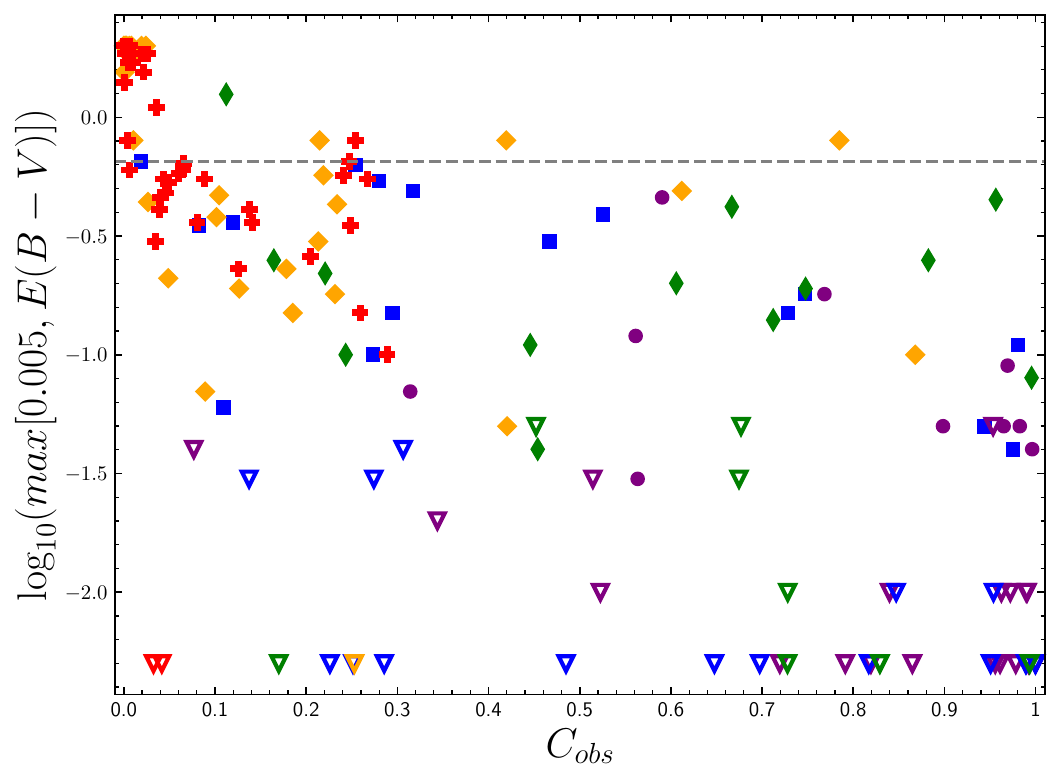}
		\caption{Optical extinction of the accretion disk component of the SED fit versus contrast. To represent $E(B-V)=0$ values in a logarithmic axis they have been approximated as $E(B-V)=0.005$. The grey, dashed line represents the $E(B-V)=0.65$ limit defined by \protect\cite{caccianiga2008}. Symbol and colour codes as in Fig. \ref{fig:lx_vs_z}, except for AGN with E(B-V) compatible with no obscuration, that have been represented with empty triangles pointing down. }
		\label{fig:EBV_Contrast}
	\end{figure}

	To look for additional insight and a potential relationship between observed contrast and extinctions, we have plotted one against the other in Fig. \ref{fig:EBV_Contrast}. It can be seen that the higher values of extinction concentrate at lower values of contrast ($C_{obs}<0.3$) while for the lowest extinction values the contrasts values show a much wider range of values. This is again related to AGN classification in the same direction as before, 1.0/1.2/1.5s with low extinctions and 1.8-9/2s being affected by high-extinction. 
	
	Despite this overall trend, there is a significant fraction of 1.8-9s and 2s with low extinction, comparable to 1.0-5s. To have a quantitative definition of what we can call high and low extinction we have used threshold value $E(B-V)=0.65$, which was found by \cite{caccianiga2008} as the amount of extinction that was able to suppress the broad component of H$\beta$ enough to be undetected. Using this value, we have 16 (50\%) 1.8-9s and 22 (50\%) 2s with low extinction, insufficient to explain their intermediate classification. We have checked that all these objects have a probability of at least 90\% to be below this limit using $\Delta \chi^2$. A similar fraction of 1.8-9 with low extinction are found in previous works in the literature \citep{Caccianiga2004,trippe2010}. For these objects some additional effect should be decreasing the observed contrast enough to explain their optical class. This will be explored in Sect. \ref{sec:lum}.
	
	We also have to take into account that the extinction measured from the SED fits accounts for the total extinction along the line-of-sight. Consequently, it could be associated to the nuclear material from the AGN, to the gas and dust of the host galaxy \citep{maiolino1995,herrero2003, baron2016,buat2021} or be a combination of both. To clarify in which situation we are, we have investigated whether there are differences in extinction associated to the host galaxy between sub-types.
	
	To estimate the extinction caused by the host galaxy we have calculated the Balmer decrement of the H$\alpha$ and H$\beta$ narrow emission lines, $F(\mathrm{H}\alpha^{\text{n}})/F(\mathrm{H}\beta^{\text{n}})$. The narrow emission lines originate far from the nuclear region of the AGN, at hundreds of parsec. Consequently, their emission is not affected by the nuclear extinction \citep{gaskell2016} but they can suffer extinction from the gas and dust of the host galaxy. The intrinsic narrow Balmer decrement distribution is not expected to change between sub-types \citep{gaskell1984}, therefore, any differences should be related to changes in the host galaxy extinction. We have considered the total flux of the narrow components of H$\beta$ and H$\alpha$ for those AGN with detection for both emission lines and both are measured in the same spectra (69\% of the working sample).
	
\begin{figure}
		\centering
		\includegraphics[width=\columnwidth]{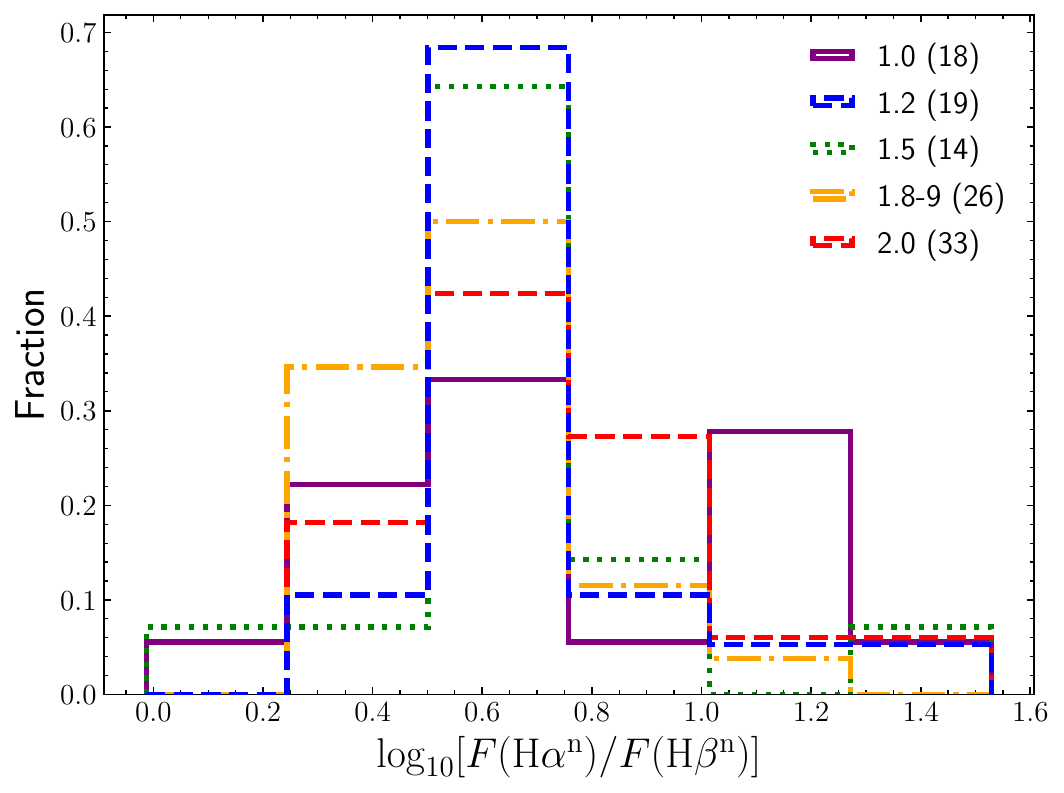}
		\caption{Distribution of the Balmer decrement of the narrow components of H$\alpha$ and H$\beta$. We indicate the number of AGN with coverage in both spectral regions. Line styles as in Fig. \ref{ec:contrast}.}
		\label{fig:BD_dist}
	\end{figure}

	The distributions of the narrow line Balmer decrement are shown in Fig. \ref{fig:BD_dist}. The distributions are visually similar for all sub-types, exhibiting a clear peak around $\mathrm{log}_{10}[(F(\mathrm{H}\alpha^{\text{n}})/F(\mathrm{H}\beta^{\text{n}})]=0.61$, comparable with previous findings in literature \citep{baron2016,lu2019,selwood2023}. The KS tests indicate the differences are not significant for any comparison between groups. If we assume an intrinsic ratio of $F(\mathrm{H}\alpha^{\text{n}})/F(\mathrm{H}\beta^{\text{n}})=3.1$ \citep{gaskell1982,gaskell1984,wysota1988}, we find that AGNs with line-of-sight $E(B-V)<0.65$ have values of extinction compatible with those derived from the narrow line Balmer decrement. For this population most, if not all, of the line of sight extinction could arise in material in the host galaxy. However, for those 1.8-9/2s with $E(B-V)>0.65$ there must be an important nuclear component contribution, which is likely playing the main role in the class determination.

	Then, according to our results, extinction is an important actor behind the contrast change and, consequently, the determination of AGN sub-type. It increases from 1.0-5 to 1.8-9/2s and follows the same trend as contrast. The increase of extinction for those 1.8-9/2s with $E(B-V)>0.65$ can be associated with a nuclear component, while for the rest, it can be explained by material in the host galaxy.

	\subsection{AGN and host galaxy luminosity}
	
	\label{sec:lum}
		
	As shown before, extinction can not be the only parameter driving the sub-type determination. A potential additional explanation would be a decrease of the AGN luminosities and/or an increase in the galaxy luminosity with sub-type, which would result in a decline in the observed contrast.
	
	Previous works have reported that type 1 AGN have higher intrinsic luminosity than type 2 AGN. This trend has been detected at different wavelengths, but mostly in X-rays \citep{dellaceca2008,burlon2011,ueda2014,lacy2015, Koss2017}. Additionally, \cite{stern2012} found a decreasing of the AGN luminosity with increasing sub-type, using H$\alpha$ luminosity as a proxy of the total luminosity. To look for any dependence with classification in our sample, we have used the extinction-corrected AGN luminosity at rest-frame 5100$\AA$, $L_{\mathrm{AGN}}$, derived from our SED fitting.
	
	To properly compare luminosities we need to account for the differences in the redshift distributions (shown in Fig. \ref{fig:lx_vs_z}). To correct for this, we assigned to each object a weight following the same procedure as in \cite{mendez2016} and \cite{mountrichas2021}. To obtain this weight, first, we obtained the distribution of redshift of the full sample and divided it in bins of $\Delta z=0.125$. Then the same process is repeated for the distribution of each sub-type. Finally, for each AGN in a particular bin $i$ of a certain sub-type, $s$, the weight, $w_{is}$, is calculated as the number of AGN in that bin, $N_i$, over the overall number of AGN of that certain sub-type in the same bin, $N_{is}$,
	
	\begin{equation}
		w_{is}=\frac{N_i}{N_{is}}.
	\end{equation}
	
	All discussion about luminosity distribution from this point will be about the redshift corrected distribution. We also accounted for the redshift correction when conducting the KS test. We weighted the luminosity cumulative distributions involved by these same weights. To implement it, we adapted of the unweighted KS test code available in the package Numpy \cite{numpy} to follow the procedure shown in \cite{monahan_2011}.
	
\begin{figure}
		\centering
		\includegraphics[width=1\columnwidth]{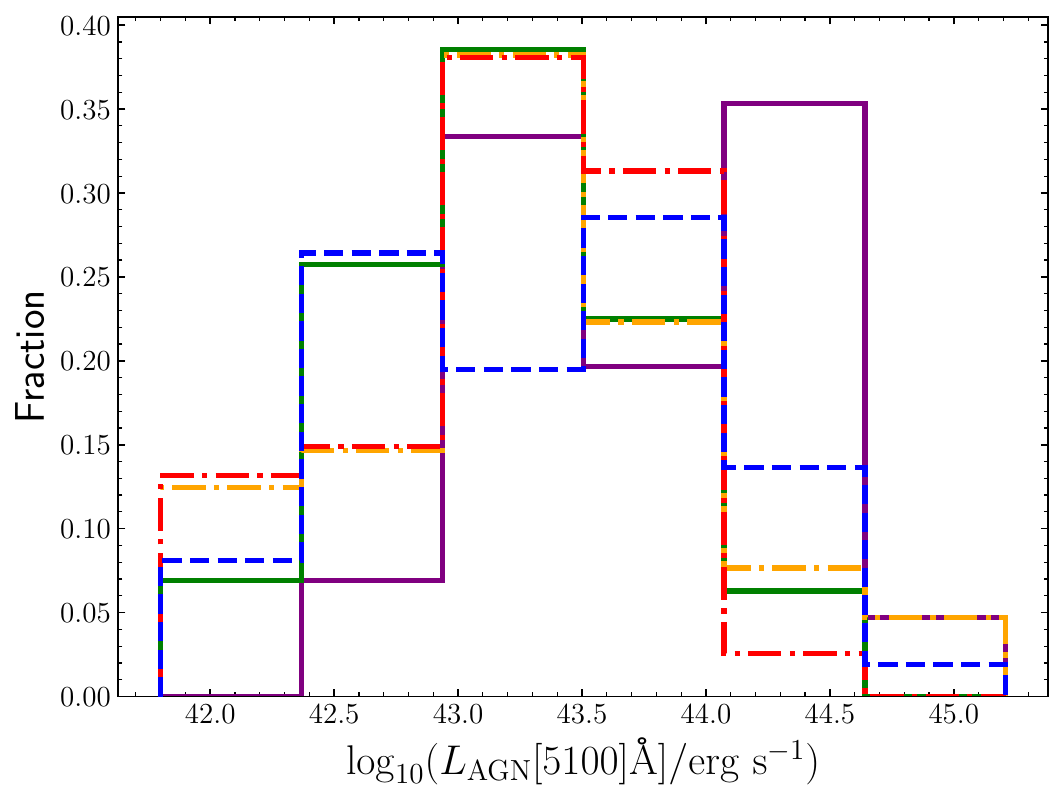}
		\caption{Redshift weighted distributions of the intrinsic AGN luminosity at rest-frame 5100$\AA$. Line styles as in Fig. \ref{ec:contrast}.}
		\label{fig:luminosityAGN}
	\end{figure}

	The distributions of $L_{\mathrm{AGN}}$ are presented in Fig. \ref{fig:luminosityAGN}. Visually, it appears that type 1.0 have slightly higher luminosities overall. In fact, the lowest values of $P(H_0)$ between individual sub types correspond to the comparison of 1.0s with 1.2s and 1.0s with 1.5s. However, no comparison between individual sub-types or even between 1.0-5 and 1.8-9/2 returns a significance greater than $3\sigma$. Similar results are found if we use instead X-ray luminosities. Based on these results, there is no evidence of a significant difference in $L_{\mathrm{AGN}}$ between sub-types, so this parameter is unlikely to play an important role in the AGN classification.

	Previous studies in the literature have found a small increase in host galaxy stellar mass for type 2 AGN, which may or may not appear statistically significant \citep{zou2019,mountrichas2021,koutoudilis2022}. The more massive host galaxies are expected to be more luminous and then the observed contrast of the hosted AGN would be smaller. In this work, we directly derived the host galaxy luminosity at rest-frame 5100$\AA$ from the SED fitting. With this approach, we are focusing on the parameter directly affecting the observed contrast.
	
\begin{figure}
		\centering
		\includegraphics[width=1\columnwidth]{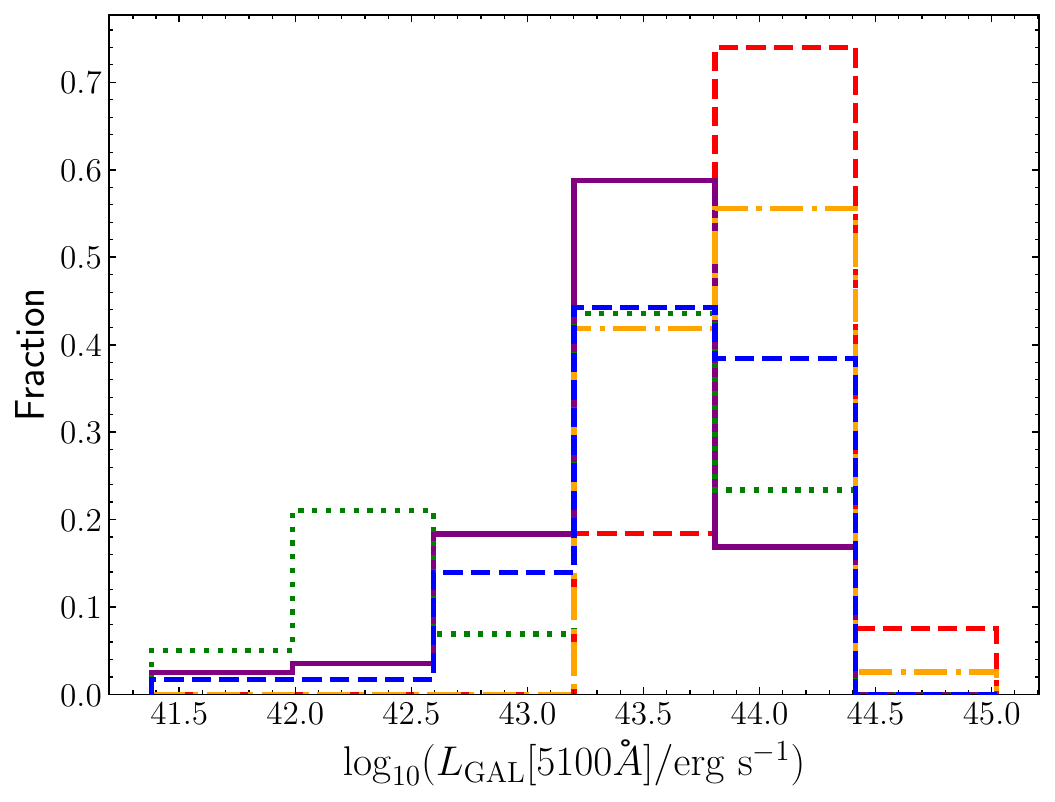}
		\caption{Redshift weighted distributions of the intrinsic luminosity of the host galaxy at rest-frame 5100$\AA$. Line styles remains as in Fig. \ref{ec:contrast}.}
		\label{fig:luminosityGal}
	\end{figure}

	We show the distributions of host galaxy luminosity, $L_{\mathrm{GAL}}$, in Fig. \ref{fig:luminosityGal}. According to the KS results, we found significant differences when comparing 1.5s with 1.8-9s and 2s, respectively (see Table \ref{tab:kstest} Col. 7). Motivated by these and previous results, we considered the comparison of 1.0-5s vs 1.8-9/2s. We found also a significant difference between the two groups. This finding is in agreement with previous results in the literature \citep{zou2019,mountrichas2021,koutoudilis2022}. As shown in Appendix \ref{sec:proyected_aperture}, the apertures of 1.0-5 and 1.8-9/2 AGN are different, so we checked if the observed difference remains if we obtain the light of the host-galaxy collected by the spectrograph. To do so, we have multiplied the host galaxy luminosity by the $f_{\mathrm{GAL}}$ obtained in Appendix \ref{sec:proyected_aperture}. The differences persist and so does the significance levels when comparing groups.
 	
\begin{figure}
		\centering
		\includegraphics[width=1\columnwidth]{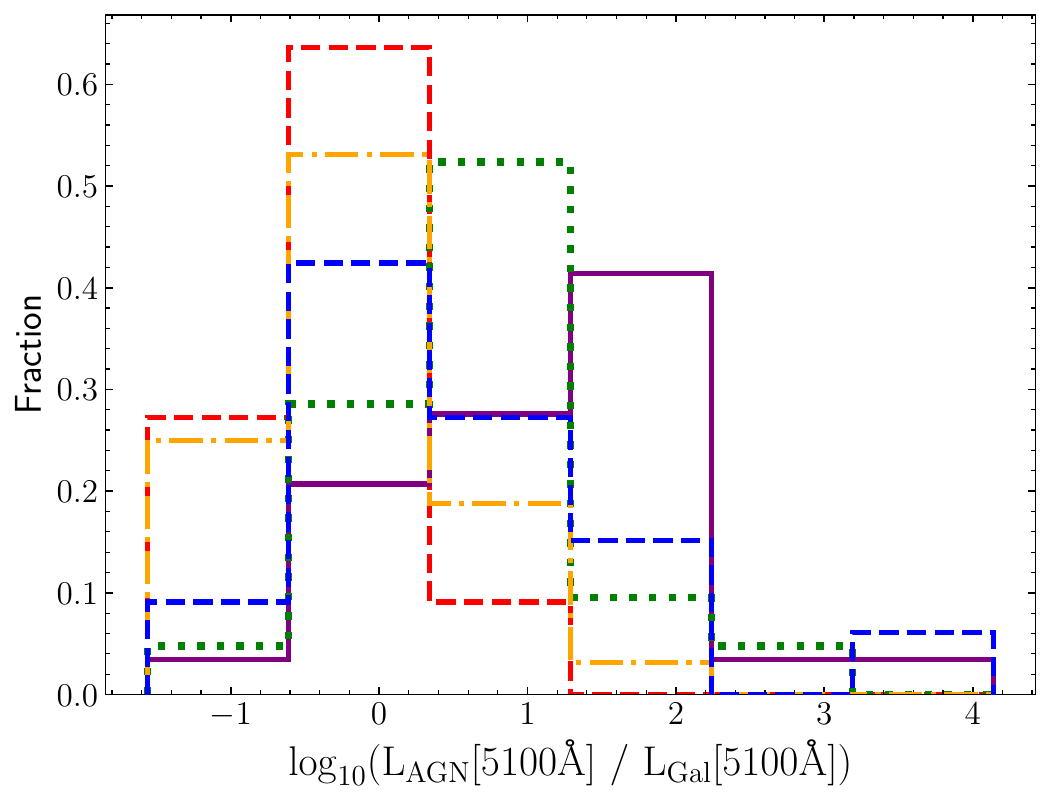}
		\caption{Intrinsic AGN over host galaxy contrast at rest-frame 5100$\AA$. Line styles as in Fig. \ref{ec:contrast}.}
		\label{fig:LAGN_LGAL_Hist}
	\end{figure}

	The observed contrast involves both the host galaxy and the AGN, and so we have also investigated their combined effect using the ratio of the intrinsic AGN over host galaxy luminosities. $L_{\mathrm{AGN}}/L_{\mathrm{GAL}}$ has the additional advantage over the individual luminosities of being a value which does not need a redshift correction. The results are presented in Fig. \ref{fig:LAGN_LGAL_Hist}. We compared the individual and grouped distributions as before. Once again, the KS test reveals a $3\sigma$ significance when comparing 1.0-5s versus 1.8-9/2s. Additionally, the combined effect of both AGN and host galaxy luminosities enhances the significance of a potential difference for several individual comparisons, although all remain below 3$\sigma$. However, the results hint at individual differences between sub-types 1.0 and 1.2-5, with similar results to those of the observed contrast distributions. We also compared the main groups limiting to $E(B-V)<0.65$ to focus on the 1.8-9/2s not explained by the extinction alone, finding also a significant difference.
	
\begin{figure}
		\centering
		\includegraphics[width=1\columnwidth]{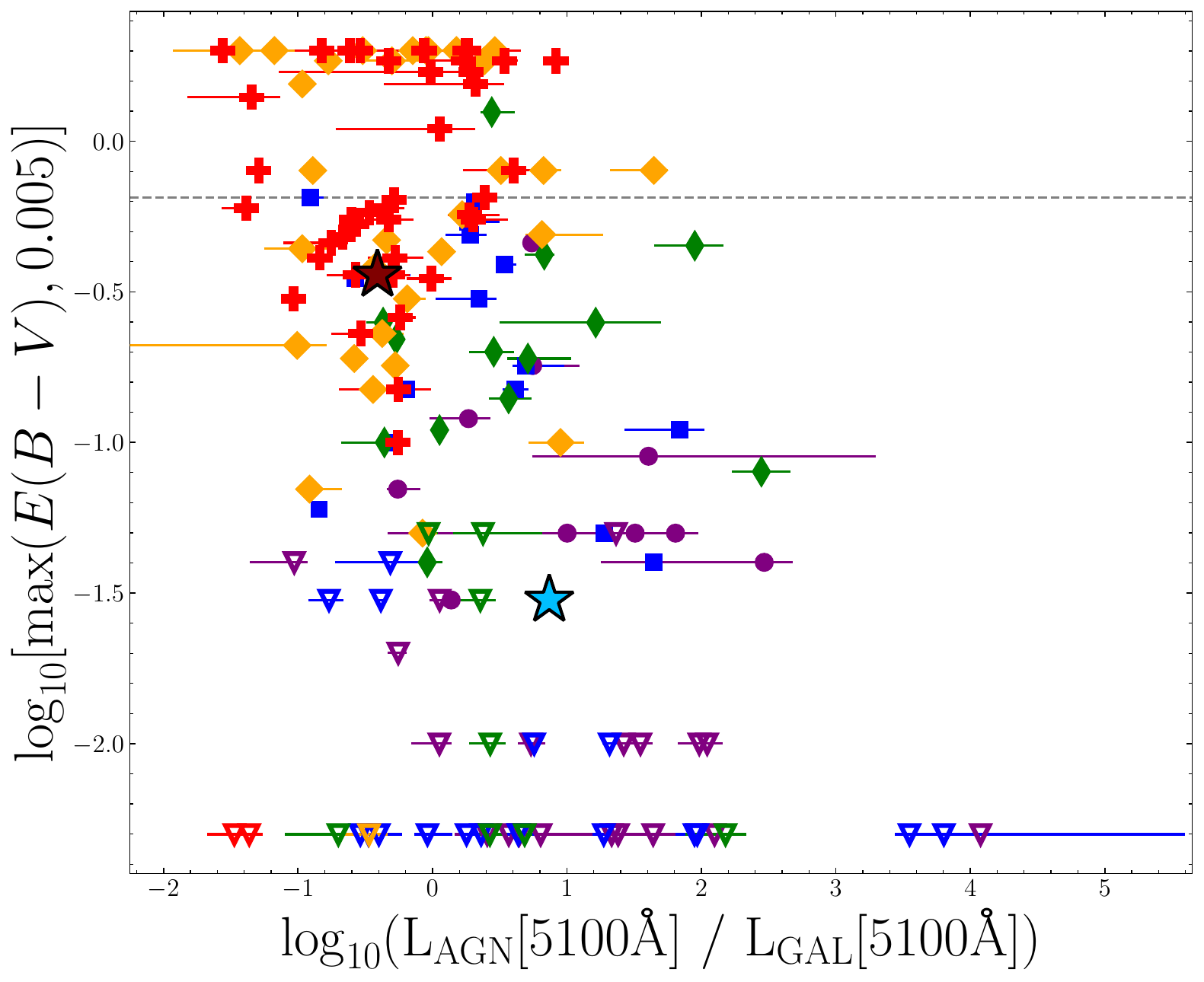}
		\caption{Intrinsic AGN over host galaxy contrast at rest-frame 5100$\AA$ versus the optical extinction. To represent $E(B-V)=0$ values in a logarithmic axis they have been approximated as $E(B-V)=0.005$. Large stars corresponds the median points of groups 1.0-5 (blue, $\mathrm{\langle \mathrm{log_{10}}(L_{\mathrm{AGN}}/L_{\mathrm{GAL}})\rangle}=0.87\pm0.11$) and 1.8-9/2 (maroon, $\mathrm{\langle \mathrm{log_{10}}(L_{\mathrm{AGN}}/L_{\mathrm{GAL}})\rangle}=-0.41\pm0.08$) both with extinctions under $E(B-V)=0.65$. Symbol and colour code as in Fig. \ref{fig:lx_vs_z}, except for AGN with E(B-V) compatible with no obscuration, that have been represented with empty triangles pointing down.}
		\label{fig:LAGN_LGAL_EBV}
	\end{figure}

	So far we have already considered the effect of the extinction and of the intrinsic luminosities separately. In order to identify the relation between both, we have represented extinction versus intrinsic luminosity ratio in Fig. \ref{fig:LAGN_LGAL_EBV}. We compared both main groups for a given extinction and found 1.8-9/2s have overall lower intrinsic AGN over host galaxy luminosity ratios than 1.0-5. In fact the mean values of $\mathrm{log_{10}}(L_{\mathrm{AGN}}/L_{\mathrm{GAL}})$ for the whole samples are $-0.23\pm0.08$ and $0.84\pm0.11$, respectively. The resultant difference is of more than one order of magnitude, $1.08\pm0.19$ dex. It becomes even larger, $1.3\pm0.2$, if we take only into account those AGN with extinctions lower than $E(B-V)=0.65$. These results suggest the following scenario: half the 1.8-9/2 objects lack enough extinction to provoke by itself their change from 1.0-5 to 1.8-9/2. In these objects a decrease of the  $L_{\mathrm{AGN}}/L_{\mathrm{GAL}}$, led mainly by an increase of $L_{\mathrm{GAL}}$, is enough to provide the extra reduction in observed contrast, necessary for that change between groups.
	
	\section{Conclusions}
	\label{sec:5}
	
	This work aimed to shed light on the effect of extinction and AGN and host galaxy luminosity on the AGN optical spectroscopic optical classification. We studied the variation of the observed contrast at rest-frame 5100$\AA$, $f_\mathrm{AGN}/(f_\mathrm{AGN}+f_\mathrm{GAL})$, extinction and luminosity of AGN and host galaxy of a large sample of 159 X-ray selected AGN with complete and robust optical spectroscopic classification. The AGNs have been drawn from the Bright Ultra-hard XMM-\textit{Newton} Survey and have $z\in[0.05-0.75]$ and intrinsic 2 -- 10 keV, absorption-corrected luminosities between $10^{42}$ and $10^{46}$ erg s$^{-1}$.
	
	To classify our AGNs we have fitted the H$\beta$, H$\alpha$ and MgII regions of their optical and/or NIR spectrum. We also have obtained the Balmer decrement of the narrow emission lines using the ratio between the narrow components of H$\beta$ and H$\alpha$. To calculate the observed AGN over total flux contrast, optical/UV line-of-sight extinction and AGN and host galaxy optical luminosities we decomposed the rest-frame UV-to-mid-infrared SED into galaxy and AGN components. The main results of our work can be summarized as follows:

	\begin{enumerate}
		\item A clear decrease of the observed AGN over total AGN+galaxy contrast with the intermediate class was found, identifying two main groups of high (1.0-1.5) and low (1.8-9/2) contrast. This is significant not only for the main groups but also for any comparison between any sub-types in the 1.0-5 and 1.8-9/2 groups. A less significant difference is hinted for sub-types 1.0 and 1.2-5.
		\item We found a similar increase in the extinction with intermediate class, showing 1.8-9/2s to have overall higher extinctions than 1.0-5s; in agreement with previous results in the literature. The shared change between contrast and extinction reveals it as one of the effects producing the change of sub-type. No significant change between sub-types inside these groups is found.
		\item We found a population of 1.8-9s (50\%) and 2s (50\%) AGN with insufficient extinction to suppress the BLR emission enough to change the classification of the AGN only by itself.
		\item The extinction in the host galaxy, gauged from the Balmer decrement of the narrow line, is not significantly different between sub-types. However, for most 1.0-5 and the low extinction 1.8-9/2.0, it could contribute significantly to the total line-of-sight extinction.
		\item No statistically significant difference is found in $L_\mathrm{AGN}$ between different AGN sub-types. If there is such a difference in our sample, it must be small and be at most a secondary effect.
		\item When comparing $L_{\mathrm{GAL}}$ between sub-types a difference is found between 1.0-5s and 1.8-9/2s and also when comparing 1.5s with 1.8-9s and 2s, with higher values for the former group in both cases.
		\item  Differences in $L_{\mathrm{GAL}}$ produce a decrease in the intrinsic ratio between AGN and host galaxy luminosity, $L_{\mathrm{AGN}}/L_{\mathrm{GAL}}$, for 1.8-9/2s. Furthermore, for a given extinction, 1.8-9/2s have lower values of $L_{\mathrm{AGN}}/L_{\mathrm{GAL}}$ than 1.0-5s.
	\end{enumerate}
	
	Based on the results presented, the sub-type classification into the main groups made of 1.0-5s and 1.8-9/2 is driven by a decreasing contrast effect. As the observed contrast decreases, the ability to detect the broad component of H$\beta$ and H$\alpha$ worsens, until is no longer possible to detect them, resulting in the change of sub-type. 
	
	By itself, extinction explains roughly the classification of 50\% of the 1.8-9/2s. For the rest, the change cannot be explained without an additional effect of a decreasing $L_{\mathrm{AGN}}/L_{\mathrm{GAL}}$ ratio (which we have found to be mainly driven by a higher $L_\mathrm{GAL}$ in 1.8-9/2.0). For these, the combination of intermediate levels of extinction and low $L_{\mathrm{AGN}}/L_{\mathrm{GAL}}$ would reduce the contrast enough to make impossible the detection of the broad components and produce the change of sub-type.
		
	Those 1.8-9/2 with extinction high enough to prevent the detection of the broad emission lines, the dominant extinction component must be associated to a nuclear contribution. However for the rest of the sample the extinction associated to the host galaxy could be a significant contribution to the total line-of-sight extinction, as shown with the estimates from the narrow line Balmer decrement estimation.
	
	We find no significant difference in any of the studied parameters between sub-types within the main groups. It is hinted a potential contrast effect for type 1.0, however, the lack of a strong significance means that the effect, in any case, would be weak. The $L_{\mathrm{AGN}}/L_{\mathrm{GAL}}$ results suggest this potential difference could be led by an increase of the $L_{\mathrm{AGN}}/L_{\mathrm{GAL}}$ in comparison to 1.2 and 1.5. Nevertheless, once again this remains only as a potential hint as we find no significant difference. Future work will focus on explaining the effect leading to the change of sub-types inside the two big groups identified using this same sample.

	\begin{acknowledgements}
		
	The authors wish to thank to Roberto Maiolino for the corrections and suggestions provided.

	LBG acknowledges economic support from the Concepción Arenal Programme of the Universidad de Cantabria. LBG, SM, FJC, AC and NC acknowledge financial support from the Grant PID2021-122955OB-C41 funded by MCIN/AEI/10.13039/501100011033 and by ERDF A way of making Europe. AAH acknowledges support from grant PID2021-124665NB-I00 funded by MCIN/AEI/10.13039/501100011033 and by ERDF A way of making Europe. IGB acknowledges support from STFC through grants ST/S000488/1 and ST/W000903/1.
	
	This work is based on observations obtained with XMM–Newton, an ESA science mission with instruments and contributions directly funded by ESA Member States and NASA. It is also based on data from the Wide-field Infrared Survey Explorer,which is a joint project of the University of California, Los Angeles and the Jet Propulsion Laboratory/California Institute of Technology, funded by the National Aeronautics and Space Administration. Funding for the SDSS and SDSS-II has been provided by the Alfred P. Sloan Foundation, the Participating Institutions, the National Science Foundation, the U.S. Department of Energy, the National Aeronautics and Space Administration, the Japanese Monbukagakusho, the Max Planck Society and the Higher Education Funding Council for England. The SDSS website is \url{http://www.sdss.org/}. This work is based on observations collected at the European Organisation for Astronomical Research in the Southern hemisphere, Chile, programme IDs 084.A-0828, 086.A-0612, 087.A-0447. It is also based on observations made with the William Herschel Telescope – operated by the Isaac Newton Group, the Telescopio Nazionale Galileo – operated by the Centro Galileo Galilei and the Gran Telescopio de Canarias installed in the Spanish Observatorio del Roque de los Muchachos of the Instituto de Astrofsica de Canarias, in the island of La Palma. This work is based on observations made with the Nordic Optical Telescope, owned in collaboration by the University of Turku and Aarhus University, and operated jointly by Aarhus University, the University of Turku and the University of Oslo, representing Denmark, Finland and Norway, the University of Iceland and Stockholm University at the Observatorio del Roque de los Muchachos, La Palma, Spain, of the Instituto de Astrofisica de Canarias. This research is based in part on data collected at the Subaru Telescope, which is operated by the National Astronomical Observatory of Japan. We are honoured and grateful for the opportunity of observing the Universe from Maunakea, which has the cultural, historical, and natural significance in Hawaii. We are grateful	to the Institute for Astronomy, University of Hawaii, for the allocated observing time on the UH88. This work is based in part on data obtained as part of the UKIRT Infrared Deep Sky Survey. This publication makes use of data products from the Two Micron All Sky Survey, which is a joint project of the University of Massachusetts and the Infrared Processing and Analysis Center/California Institute of Technology, funded by the National Aeronautics and Space Administration and the National Science Foundation.
	
	In this work we have used the python packages ASTROPY \citep{,astropy:2022}, Matplotlib \citep{matptlolib}, Numpy \citep{numpy} and Scipy \citep{scipy} for the analysis and representation of our data. For the initial exploration of our data we used the software TOPCAT \citep{topcat}. The reduction of spectra in this work has been carried out using IRAF \citep{doug1986} and PyEmir \citep{pascual2010,cardiel2019}. This research has made use of software provided by the Chandra X-ray Center (CXC) in the application packages CIAO and Sherpa \citep{sherpa}. For the SED fitting we have used the SEd Analysis using BAyesian Statistics (SEABAS; \citealt{rovilos2014}) program. 
	
	The optical and NIR fitting results of the BUXS sample will be made	publicly available in Mateos et al. (In preparation). The other data products underlying this article will be shared upon a reasonable request to Silvia Mateos.
	
	\end{acknowledgements}


\bibliographystyle{aa}
\bibliography{referencias}

\begin{appendix}
	
	\section{Fitting of continuum and emission lines fitting}
	
	\label{sec:apendix_emission_lines}
	
	To obtain the optical classification and emission line properties, we needed to fit the emission lines and continuum around MgII, H$\beta$ and H$\alpha$, respectively. In this appendix, we explain the general procedure and the models used for the continuum and emission lines.
	
	To obtain the best constraints, we fitted H$\beta$ and H$\alpha$ independently. We have first fitted the local continuum in a region free of emission lines defined around H$\beta$ and H$\alpha$. We considered the same wavelength regions that \cite{shen2011} used for both the continuum and the emission lines. If a part of the regions was affected by atmospheric absorption, we removed it from the fit. We provide the values used to define every region in the corresponding subsections below. 
	
	To avoid over-fitting and adding non-necessary components, we increased gradually the complexity of the emission line model. We accepted a more complex model if an F-test returned a 5$\sigma$ improvement after including an extra component or parameter. 
	
	For all AGN we had an initial redshift estimation from a visual fit of the position of at least one of the Balmer emission lines or the [OIII] doublet. The redshift was then let to vary while fitting to obtain a more accurate value. For 2 type 2 AGN there was no clear detection of the prominent emission lines, even though their nature as AGN was confirmed using the X-ray information. In these AGN, the initial redshift estimation was obtained from the absorption features of CaII as their continuum was dominated by the host galaxy. 
	
	We corrected the intrinsic width of all emission lines fitted for the instrumental broadening of the spectrographs employed. The value of the instrumental broadening was directly available for SDSS spectra. For the rest, we fitted of the atmospheric lines of the sky spectrum to derive it.
	
	\subsubsection{Iron Emission Template}
	
	AGN show a pseudo-continuum FeII emission across the UV and optical ranges. This emission is a consequence of the blending of a large number of individual emission lines, produced by the high-velocity movement of the gas in which they originated. In some AGN, the FeII emission can contribute significantly to the continuum in the H$\beta$ and MgII ranges, so it is fundamental to account for this component.
	
	To model the FeII emission we used the theoretical prediction from \cite{verner2009}. It covers from $1000$ $\AA$ to $10\,000$ $\AA$ and was built using an 830-level model atom for FeII. There are different templates available that differ in the $N_H$, the microturbulence and the ionizing flux. Here we used the one that best fits I ZW 1, the prototypical FeII emission of AGN, as used in other works in the literature \citep{dong2008,jin2012,calderone2017}.
	
	The template has three degrees of freedom: the normalisation, a wavelength shift up to $\pm 1800\ \mathrm{km}\ \mathrm{s}^{-1}$ and the width of a Gaussian function used to convolve with the iron template to model the blending of the FeII emission lines, up to $10000\ \mathrm{km}\ \mathrm{s}^{-1}$.
	
	\subsubsection{Galaxy Template}
	
	Previous works have shown that luminous AGN, as the ones in our sample, are most likely hosted by elliptical galaxies \citep{floyd2004,polleta2007,calderone2017,coffey2019}. Consequently, we have used the composite spectrum obtained for $\sim$2000 early-type galaxies from SDSS DR7\footnote{Template number 23 in \url{http://classic.sdss.org/dr7/algorithms/spectemplates/}} to fit the host galaxy contribution. When fitting the galaxy template we have two free parameters: normalisation of the template and a small wavelength shift up to $\pm 600\ \mathrm{km}\ \mathrm{s}^{-1}$ associated with small corrections to the input redshift. 
	
	As this template was obtained from a combination of spectra measured with the SDSS spectrograph, we need to correct it for the difference in spectral resolution with other spectrographs used for a fraction of the spectra in our working sample (see Sect. \ref{sec:spectra}). First, we deconvolved the host galaxy template with a Gaussian with width equal to the mean instrumental width of the SDSS spectra in our sample. Then, we convolved the resultant spectrum with a Gaussian with a width equal to that of the instrumental width obtained from the sky spectrum associated with each observation.
	
	\subsubsection{Disk Continuum Model}
	
	As in previous works in the literature \citep{shen2011,Koss2017,coffey2019} we used a local power-law to model the AGN continuum associated with the disk. The power law was defined as,
	
	\begin{equation}
		f_\lambda=A\left( \frac{\lambda}{5100\AA} \right)^\alpha.
	\end{equation}
	
	The index of the power law, $\alpha$, was limited to take values in the range $[-6,\ 6]$ and the normalisation factor, $A$, was left free to take any positive value and corresponds to the $f_\lambda$ value at 5100\AA. As the disk continuum is only fitted in the region around H$\alpha$, H$\beta$ and MgII, and for obscured AGN the continuum will be dominated by the host galaxy light, we will not model the associated extinction, as it would be poorly constrained.
	
	\subsubsection{MgII}
	
	Although we did not used the properties of the MgII emission lines in this work, their detection was employed to identify the AGNs classified as type 1.x and 2.x. And so, for completeness, the description of the procedure to fit them is included here.
	
	The region of MgII emission, as defined by \cite{shen2011}, covers the wavelengths [2700, 2900] \AA. The continuum fitting regions were [2200, 2700] \AA\ and [2900,3090] \AA. The continuum was modelled locally using a combination of the accretion disk power law and the iron template (see above). We have coverage of MgII for 41 AGN ($\sim 26\%$ of the sample).
	
	We fitted two models: one model with only a continuum and a narrow component of MgII and the other one with the addition of a broad emission line. Although MgII is a doublet, the proximity of the two lines, combined with significant line broadening, causes it to be detected as a single line. We described the narrow and broad components using one Gaussian component for each one. We required the broad emission line to have a width of at least 1000 km s$^{-1}$ and no bigger than 20000 km s$^{-1}$. The narrow line was required to be narrower than 2000 km s$^{-1}$. 
	
\begin{figure}
		\centering
		\includegraphics[width=\columnwidth]{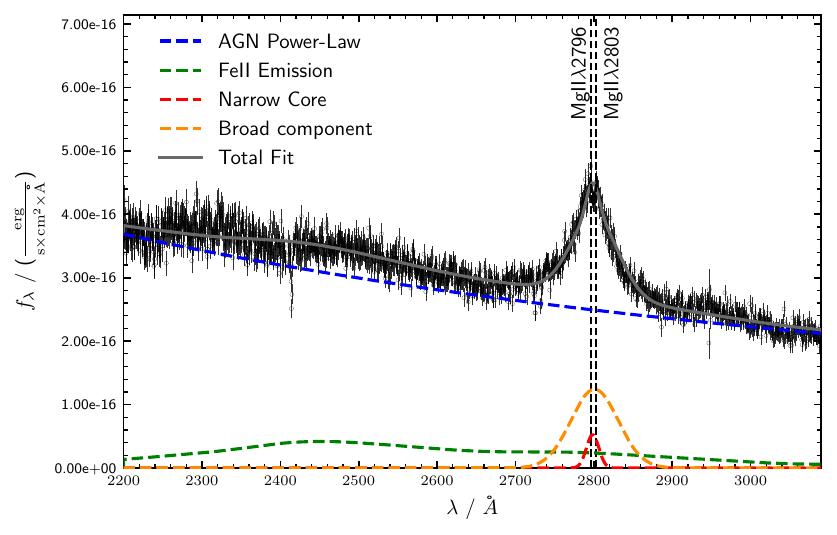}
		\caption{Example of a spectral fitting in the MgII region. The spectrum corresponds to a type 1.0 AGN. Wavelength is represented in rest-frame.}
		\label{fig:mgii_example}
	\end{figure}

	The overlapping width ranges allowed us to fit cases with exceptionally wide narrow emission lines ($>1000\ \mathrm{km\ s^{-1}}$) or exceptionally narrow broad emission lines ($<2000\ \mathrm{km\ s^{-1}}$). However, for no AGN the narrow emission line was wider than the broad component. One example of a fit in the MgII region can be seen in Fig. \ref{fig:mgii_example}.
	
	\subsubsection{H$\beta$ and [OIII]}
	
	We define the region of H$\beta$ emission line as the wavelengths in the range [4700,5100] \AA. The continuum fitting regions were [4435, 4700] \AA\ and [5100, 5535] \AA.  We modelled the continuum using a combination of the host galaxy template, the accretion disk power law and the iron template.
	
	Each one of the narrow emission lines was fitted with one Gaussian with separation fixed by the theoretical centroids of [OIII]$\lambdaup\lambdaup$4959,5007\AA \ and H$\beta$. The ratio between the [OIII] has been fixed to the theoretical value of 2.98 \citep{storey2000} and the three lines share the same width that can go up to 2000 km s$^{-1}$. The narrow emission lines could also be displaced by small wavelength shift up to $\pm 600\ \mathrm{km}\ s^{-1}$ to take into small corrections to the input redshift, similar to the host galaxy template.
	
	The narrow emission lines [OIII]$\lambdaup\lambdaup$4959,5007 can show blue wings or other complex profiles \citep{boroson2005,shen2011,rojas2020}. To properly fit them, we added a Gaussian component with the same constraints in their relative wavelengths and fluxes but different widths, up to 2500 km s$^{-1}$. The wings of the narrow emission line could also be shifted up to $\pm 1800\ \mathrm{km}\ s^{-1}$, to account for the wavelength shift relative to the core components.
	
	For the broad component of H$\beta$, we initially used one Gaussian. However, for many AGN a single Gaussian is not enough to correctly fit its profile. To ensure a good fit, we consider an additional Gaussian component if it improves the fit by $5\sigma$. Both components are required to have widths greater than 1000 km s$^{-1}$ and limited up to 20000 km s$^{-1}$. As for MgII, although the width ranges are overlapping, no fit returned a narrow emission line wider than the broad component. Both components could have a wavelength shift up to $\pm3500$ km s$^{-1}$, to allow the fit of complex profiles. 
	
	After the visual screening, we rejected the additional component when it did not describe the broad component. It can happen in several ways. In some cases, the additional Gaussian component has a width on the upper-limit range of the allowed range of widths. Usually, when this happens the Gaussian used to describe the broad component is instead misfitted as an additional component to fit the continuum. Sometimes the opposite situation happens, a line width in the lower limit of possible values. In these cases, the broad component is misused to fit small features close to the central wavelength of H$\beta$. It can also happen that the displacement of the broad component goes to the limits of $\pm3500$ km s$^{-1}$ and the program misuses the component fitting a bump unrelated to the broad component of H$\beta$, associated with the shape of the observed local continuum. We identified all these non-physical artefacts after a visual analysis. We rejected the resulting fit and forced a new one without this additional component, even if the former combination was formally improving the fit by $5\sigma$. 
	
\begin{figure}
		\centering
		\includegraphics[width=\columnwidth]{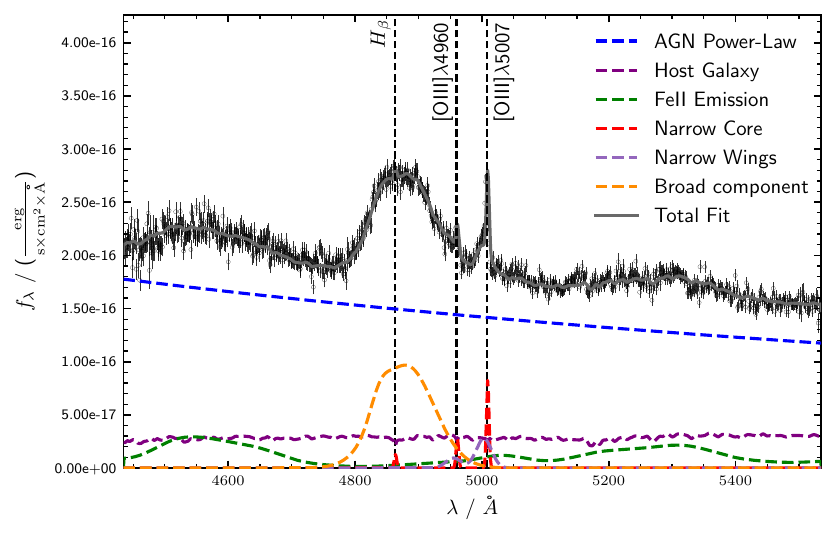}
		\caption{Example of the spectral fitting in the H$\beta$ region. The AGN dominates the continuum and a strong broad component of H$\beta$ modelled with two Gaussian components is detected. The additional component of the narrow emission line is labelled as 'Narrow Wings'. The plot corresponds to a 1.0 AGN. Wavelength is represented in rest-frame.}
		\label{fig:hb_example}
	\end{figure}

\begin{figure}
		\centering
		\includegraphics[width=\columnwidth]{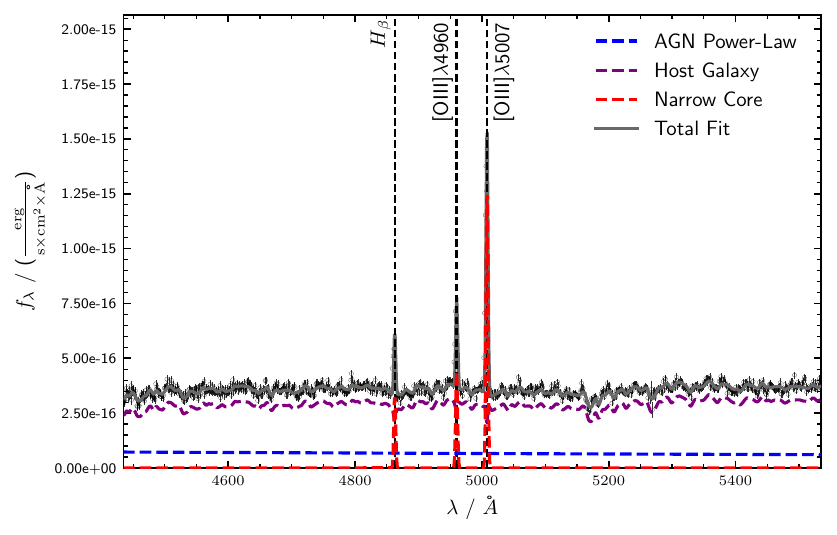}
		\caption{Example of the spectral fitting in the H$\beta$ region with a continuum dominated by the host galaxy. The plot corresponds to a type 2 AGN. Wavelength is represented in rest-frame.}
		\label{fig:hb_example_2}
	\end{figure}

	Two examples of fits in the region of H$\beta$ with and without a broad component fitted can be seen in Figures \ref{fig:hb_example} and \ref{fig:hb_example_2}.

	\subsubsection{H$\alpha$, [NII] and [SII]}
	
	We defined the region of the H$\alpha$ emission line as the wavelengths in the range [6400,6800]\ \AA. The continuum fitting regions were [6000, 6250]\ \AA\ and [6800, 7000]\ \AA. The continuum was modelled locally in the same way as in the $H\beta$ region, but without adding a FeII contribution, as its contribution is negligible at these wavelengths.
	
\begin{figure}
		\centering
		\includegraphics[width=\columnwidth]{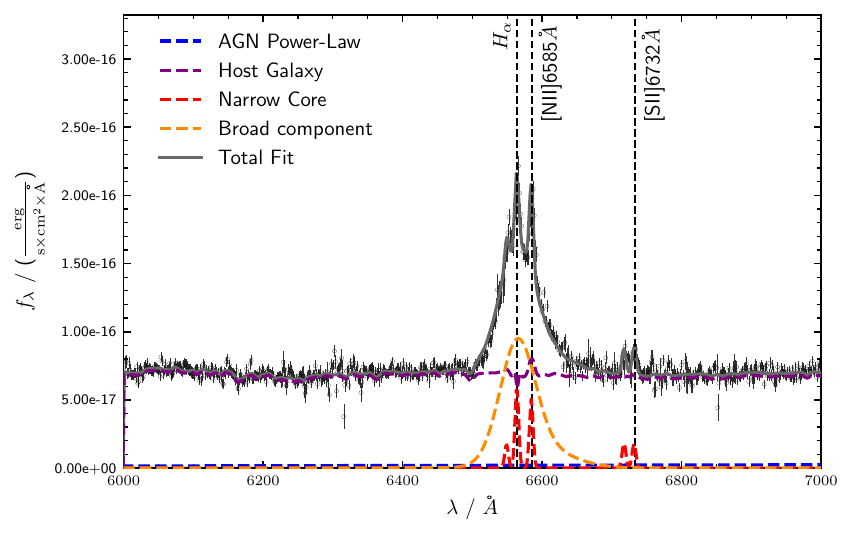}
		\caption{Example of the spectral fitting in the H$\alpha$ region with broad H$\alpha$ component. It corresponds to a 1.2 AGN. Wavelength is represented in rest-frame.}
		\label{fig:ha_example}
	\end{figure}

\begin{figure}
		\centering
		\includegraphics[width=\columnwidth]{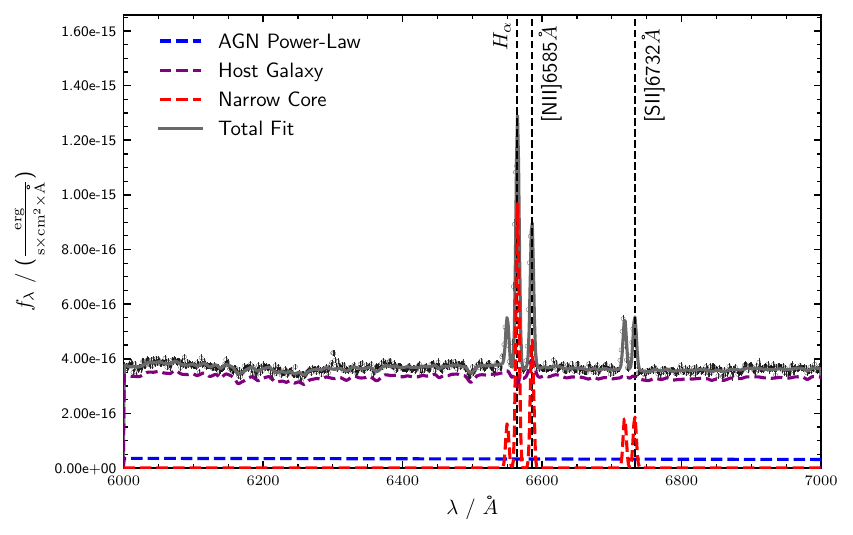}
		\caption{Example of the spectral fitting in the H$\alpha$ region without broad H$\alpha$ component. It corresponds to a type 2 AGN. Wavelength is represented in rest-frame.}
		\label{fig:ha_example_2}
	\end{figure}

	To fit the emission lines in the region of the H$\alpha$ (H$\alpha$, [NII] and [SII]) we used the same approach as for the H$\beta$ region. However, as the narrow emission lines [NII] and H$\alpha$ are heavily blended, the determination of the wings is problematic. The blending of the wings of the narrow components could be misinterpreted as a broad component, as shown in \cite{shimizu2018}. To overcome these problems, we added the wing components to the narrow emission lines only if they had been previously detected in the H$\beta$ region. In this case, we forced them to have the same width and wavelength shift as the wings in H$\beta$. Two examples of fits in the region of H$\alpha$ with and without a broad	component can be seen in Fig. \ref{fig:ha_example} and Fig. \ref{fig:ha_example_2}.
	
	As the broad components of H$\alpha$ are brighter than those of H$\beta$, they are usually better constrained. To confirm the goodness of the fit of H$\beta$ and reject non-physical fits, we compared their flux, width and displacement relative to the expected central wavelength. If for two parameters we found inconsistencies, we rejected the detection of broad H$\beta$.
	
	To compare the flux of the broad H$\alpha$ and H$\beta$ components we used the Balmer decrement. Although the intrinsic value is still a debated topic in literature  \citep{dong2008,gaskell2017,lu2019}, one of the lowest central values of its intrinsic distribution proposed in the literature is $F({\rm H}\alpha^{\rm b})/F({\rm H}\beta^{\rm b})=2.7$ \cite{gaskell2017}. This value can increase with extinction, but lower values can be only due to the dispersion of the intrinsic distribution, so, we can not expect much more bluer values. Any AGN with a Balmer Decrement lower than 2.5 (lower $5\sigma$ range of the intrinsic distribution of \cite{gaskell2017}) was marked as inconsistent according to the flux of the broad components. For the comparison of widths, we considered that H$\beta$ is systematically broader than H$\alpha$ following the relation of \cite{greene2005}. Using this method we were able to correct a total of 18 spurious broad H$\beta$ fits.
	
	\section{Projected aperture effect in class determination}
	
	\label{sec:proyected_aperture}
	
	The classification of our sources, as defined in Sect. \ref{sec:classification}, may be dependent on the projected aperture. The fraction of the host galaxy light collected by the spectrograph increases with the projected aperture because of the extended nature of the host galaxy ($\sim$ several kiloparsecs) while the unresolved emission from the AGN ($\sim$ hundreds of parsecs) will remain constant.
		
	We would then expect an increasing dilution of the AGN light by the host galaxy with the projected aperture. The increased host-galaxy contribution will hamper the detection of broad emission lines and increase the measured [OIII]5007$\AA$ luminosity. The former would favour the identification of 1.9 and 2 AGN at larger projected apertures. The latter would produce a decrease of $R$ with the projected aperture, changing the classification $1.0\rightarrow1.2\rightarrow1.5$ with increasing projected aperture. 
		
	In this appendix, we briefly study both effects to determine if we have a significant aperture effect in our class determination. To do so we obtained the projected aperture for all of the objects from their redshift and the size of the slit or fibre used in the observation. First, we calculate the angular distance, in kpc/$''$, of our objects from their redshifts applying the cosmology described at the end of Sect. \ref{sec:introduction}. We then obtained the projected aperture by multiplying it by the aperture of the slit or fibre in arcseconds.
		
	To carry out our analysis, we also calculated the fraction of host-galaxy light inside the projected aperture. We have modelled the surface brightness of the host galaxy as a De Vaucouleurs profile with an $R_e=3\ \mathrm{kpc}$, the typical value for early-type galaxies in our range of luminosities \cite{shen2003}. When integrating the profile for a fibre we used a radial integral using half the projected aperture as the integration radius. For a slit-based observation, we carried out a 2D integration of a square area. The size of one side of the square was the projected aperture while for the other we obtained the projected aperture for $2''$, the extraction diameter defined in the reduction of the spectra.
	
\begin{figure}[h]
		\centering
		\includegraphics[width=1\columnwidth]{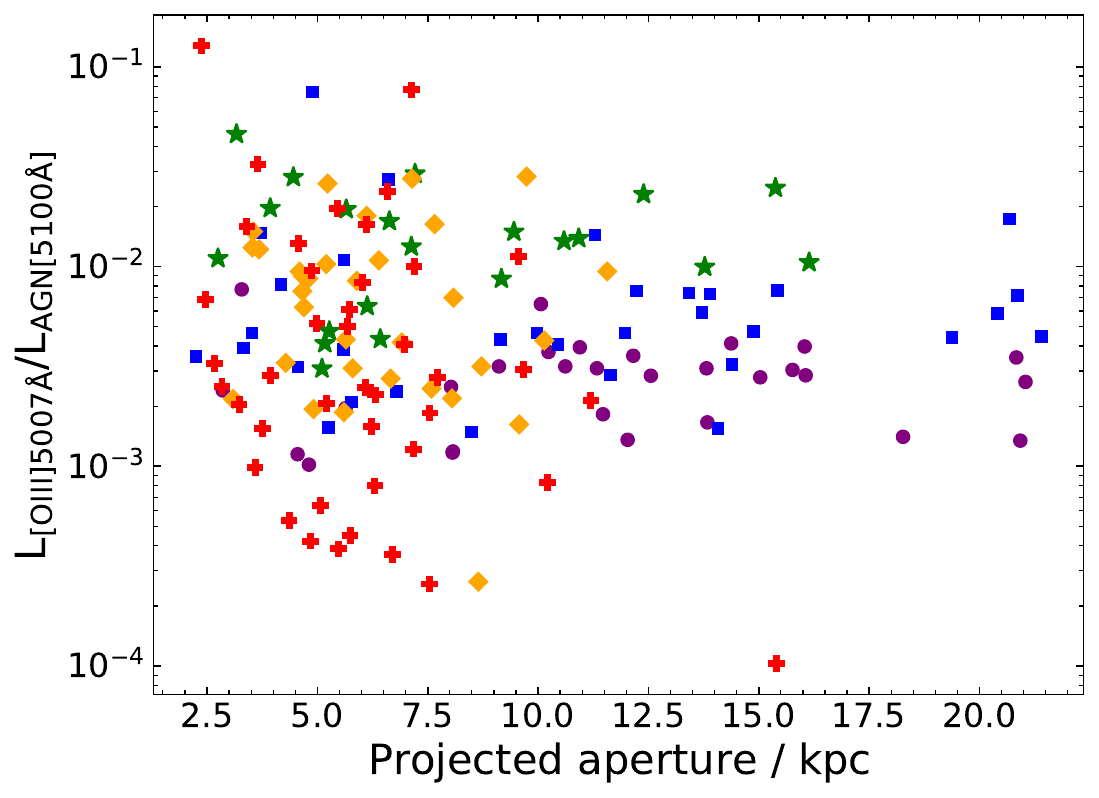}
		\caption{Intrinsic AGN over [OIII]5007\AA\ luminosity ratio versus the projected aperture for all AGN in the sample. Symbol and colour codes as in Fig. \ref{fig:lx_vs_z}.}
		\label{fig:LOIII_aperture}
	\end{figure}

	We first estimated the increase of the host galaxy contribution to the measured [OIII]5007\AA\ luminosity. To measure its strength in relation to the AGN, we obtained the luminosity ratio between AGN at rest-frame 5100\AA\ and the [OIII]5007\AA \ emission line. We then compared it against the projected aperture, as shown in Fig. \ref{fig:LOIII_aperture}. The former ratio is equivalent to the EW of [OIII]5007\AA . Any significant host galaxy contribution associated with an aperture effect should produce an increase of such ratio with the projected aperture. The lack of any significant correlation is confirmed with a $\tau$-test ($P(H_0)\sim0.29$).

\begin{figure}
		\centering
		\includegraphics[width=1\columnwidth]{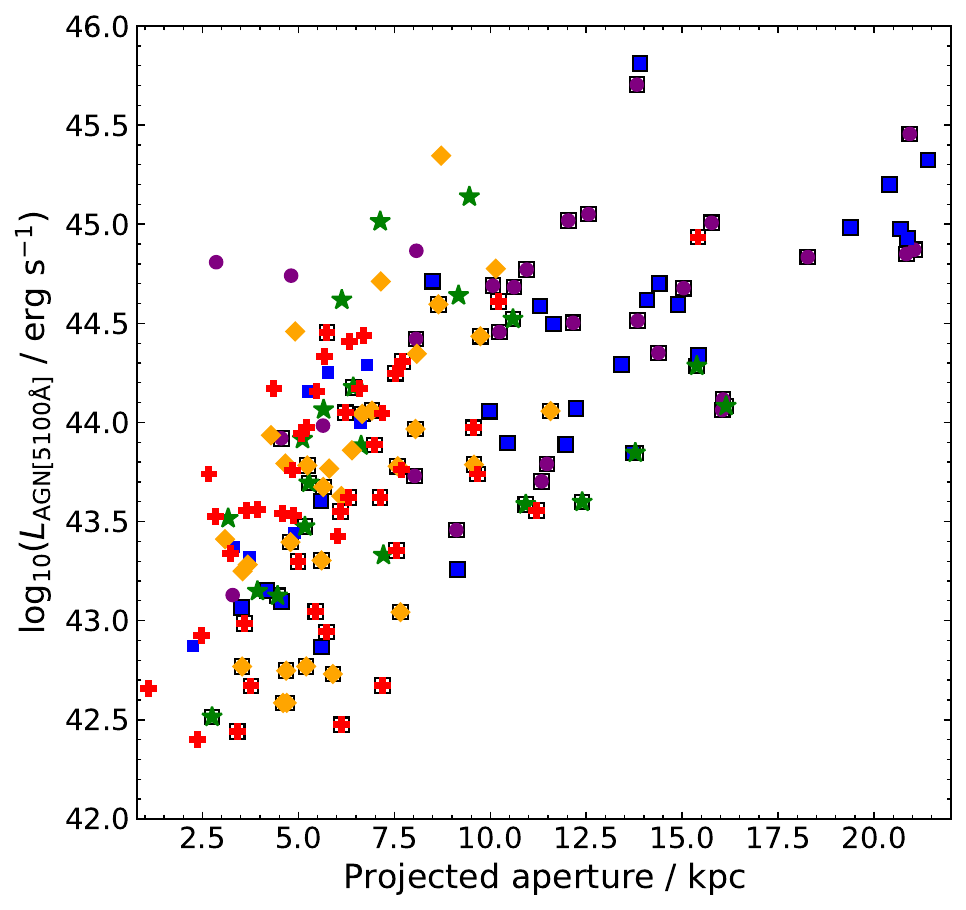}
		\caption{Decimal logarithm of the intrinsic AGN luminosity at rest-frame 5100\AA\ ratio versus the projected aperture. Sources with fibre spectrum from SDSS are marked with empty dark squares. Symbol and colour codes as in Fig. \ref{fig:lx_vs_z}}
		\label{fig:lagn_aperture}
	\end{figure}

\begin{figure}
		\centering
		\includegraphics[width=1\columnwidth]{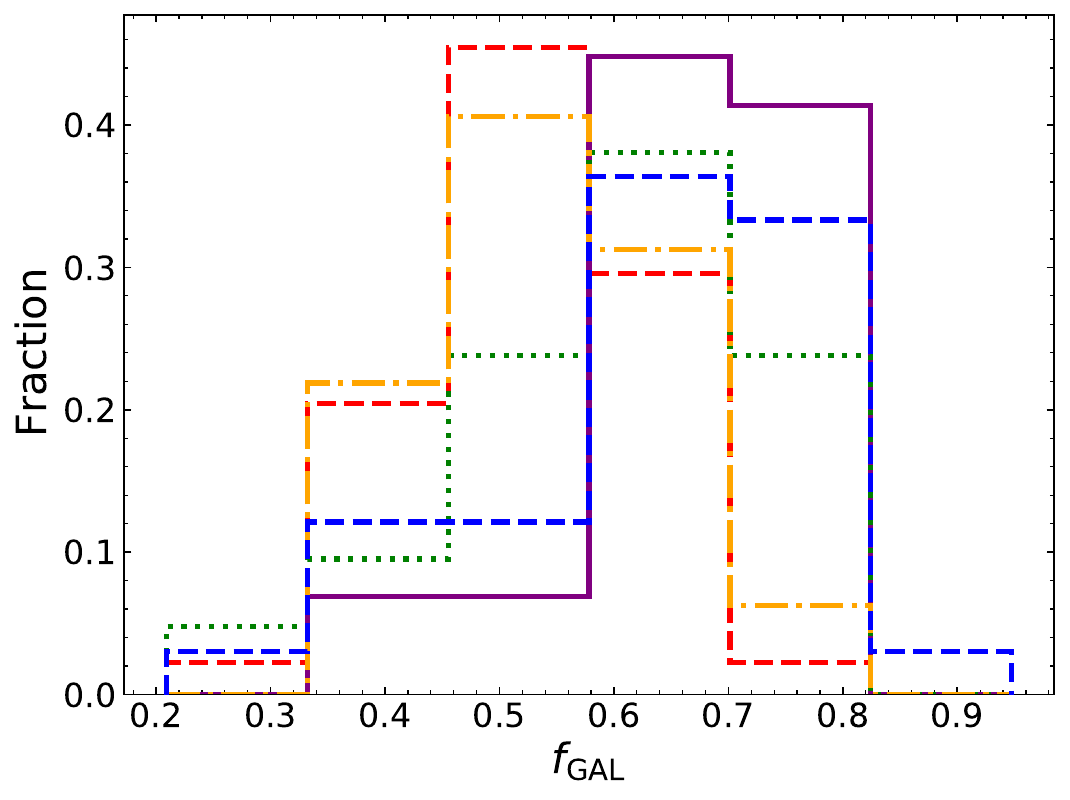}
		\caption{Distributions of the fraction of light of a galaxy with a De Vaucouleurs profile and $R_e=3\ \mathrm{kpc}$ collected by the instrumental configuration of every AGN grouped by subtypes. Line styles remains as in Fig. \ref{ec:contrast}.}
		\label{fig:fraction_galx}
	\end{figure}

	To check for the potential effect of host galaxy dilution of the broad emission lines we have followed two approaches. First, we compared the distribution by subtypes of our sample in an AGN luminosity versus projected aperture diagram (Fig. \ref{fig:lagn_aperture}). Second, we compared the fraction of host-galaxy light inside the projected aperture by subtypes (Fig. \ref{fig:fraction_galx}). We would expect that an aperture effect would place 1.9/2 at higher projected apertures for a given luminosity and at higher fractions of host-galaxy light observed. If we don't observe either we can confidently affirm that the effect of aperture in our class determination would be negligible.
	
	In Fig. \ref{fig:lagn_aperture} we compare the AGN luminosity at rest-frame 5100\AA\ versus the projected aperture. 1.9/2 are not at higher projected apertures but instead, they are preferentially at the lowest apertures. This is due to a selection effect. To carry out the spectroscopic follow-up of our AGN sample, we first collected all available SDSS spectra and then carried out a follow-up of those sources without them. We used long-slit observations with typical sizes of $\sim 1''$ so the projected apertures are smaller than those of the $2''/3''$ fibre-based SDSS spectra. The SDSS spectroscopic survey is biased towards blue type 1 AGN and selected preferentially 1.0-5s. As such, identification of 1.8-9/2s has been carried out mostly with our observations which explains the position of 1.0-5 in the diagram.
	
	We have also checked for differences in the fraction of host-galaxy light in the spectra, $f_{\mathrm{GAL}}$, by sub-types.  We show the distributions in Fig. \ref{fig:fraction_galx}. As can be seen, 1.9 and 2 AGN have lower $f_{\mathrm{GAL}}$ overall.
	
	The combined evidence of both analyses is enough to conclude that any effect of dilution by aperture is not significant in the class determination of our sample.
	
\end{appendix}	

\end{document}